\documentclass[pdflatex,sn-mathphys-num]{sn-jnl}% Math and Physical Sciences Numbered Reference Style
\usepackage{graphicx}%
\usepackage{multirow}%
\usepackage{amsmath,amssymb,amsfonts}%
\usepackage{amsthm}%
\usepackage{bm}
\usepackage{mathrsfs}%
\usepackage[title]{appendix}%
\usepackage{xcolor}%
\usepackage{textcomp}%
\usepackage{manyfoot}%
\usepackage{booktabs}%
\usepackage{algorithm}%
\usepackage{algorithmicx}%
\usepackage{algpseudocode}%
\usepackage{listings}%
\usepackage[subrefformat=parens]{subcaption}

\usepackage{xcolor}
\usepackage{soul}
\theoremstyle{thmstyleone}%
\theoremstyle{thmstyletwo}%

\theoremstyle{thmstylethree}%

\begin{document}

\title[Article Title]{Minimal seed in supersonic boundary layer at $M=3$}


\author*[1]{Nobutaka Taniguchi}\email{nobutaka.taniguchi.c1@tohoku.ac.jp}

\author[1]{Aiko Yakeno}\email{aiko.yakeno@tohoku.ac.jp}
\equalcont{These authors contributed equally to this work.}

\affil*[1]{\orgdiv{Institute of Fluid Science}, \orgname{Tohoku University}, \orgaddress{\street{2-1-1 Katahira}, \city{Sendai}, \postcode{9808577}, \state{Miyagi}, \country{Japan}}}


\abstract{This study investigates the bypass transition process in supersonic boundary layers at $M=3.0$ and $Re=300$ using adjoint-based nonlinear non-modal analysis. We demonstrate that the full nonlinearity induces a rapid path for laminar-to-turbulent transition distinct from the linear amplification mechanism. The identified nonlinear optimal perturbation (NLOP) is characterized by its locally staggered vortex structures distinct from linear solution, which promotes nonlinear interactions in wall-normal direction. Based on the investigation of its temporal evolution, a key finding is that the nonlinear amplification mechanism induces a spatially-twisted S-shaped vortex structure, causing local breakdown to trigger the transition. Furthermore, we find that the thinning of boundary layer under the cooled-wall condition promotes these nonlinear interactions, despite the linear stabilizing effect of wall-cooling.}




\maketitle

\section{Introduction}\label{sec1}

The fundamental properties of compressible boundary layers have been studied for decades. In linear stability theory (LST), whether the relative Mach number of disturbances to uniform flow is entirely subsonic or partially supersonic determines their linear characteristics \cite{lees1946investigation, mack1984boundary}. In inviscid and local theory, the existence of a generalized inflection point inside the compressible boundary layers was proved as the necessary condition for neutral subsonic disturbances in the pioneering study of Lees and Lin (1946)~\cite{lees1946investigation}. The numerical results of these inviscid neutral modes were examined and reviewed in \cite{mach1975linear, mack1984boundary}, including the first mode and higher modes. Contrary to the incompressible results, the first mode is most unstable for oblique waves, and cooling the boundary wall suppresses this instability by eliminating the generalized inflection point~\cite{mack1984boundary}. The effects of viscosity were investigated using asymptotic expansions in \cite{lees1946investigation, lees1962stability} and in the full Navier--Stokes (N--S) equations by \cite{mack1984boundary, malik1990numerical}. For the first mode, the viscous effect and non-parallel effect have been incorporated in the method of matched asymptotic expansions for compressible boundary layers by Smith (1989) \cite{smith1989first}. He demonstrated that the growth rate of the first mode deviates from parallel-flow approximation for high Mach number and large oblique wave angles. 

These linear instabilities play an essential role in the transition path, especially for small-amplitude disturbances. The dominant transition mechanisms in compressible boundary layers with various disturbance environments are summarized in \cite{morkovin1985bypass, reshotko2008transition, fedorov2011transition}. A low supersonic flat-plate boundary layer undergoes transition due to the secondary instability including oblique wave breakdown and subharmonic transition \cite{ng1992secondary, chang1994oblique, adams1996subharmonic}. Chang and Malik (1994)~\cite{chang1994oblique} applied the nonlinear parabolized stability equations (NPSE) to a supersonic boundary layer, and they found that the wave-vortex interaction produces a rapid breakdown based on a nonlinear interaction between a pair of oblique first mode waves and streamwise steady vortex. This transition process has also been reported in incompressible boundary layers~\cite{berlin1999numerical}. Regarding the effect of spanwise vortices, Adams and Kleiser (1996)~\cite{adams1996subharmonic} analyzed the evolution of slightly perturbed spanwise vortices near both the critical layer and wall at Mach number $M=4.5$ with a deformation of upper vortices into S-shape. The existence of these various transition paths supports the importance of identifying the most rapid mechanism of the disturbance growth. 

More recently, non-modal analysis has seen continuous development in compressible flow fields. Non-modal analysis was initiated by the finding that the specific combination of eigenmodes in LST had the potential to cause rapid disturbance growth due to the non-normality of linearized N--S operator~\cite{trefethen1993hydrodynamic, reddy1993energy}. Hanifi et al. (1996)~\cite{hanifi1996transient} formed a basic understanding of linear transient growth in parallel-flow approximated compressible boundary layers, demonstrating that the flow sustains the lift-up mechanism and $Re$-scaling in the compressible case, as found in incompressible flows. This temporal theory was extended to the spatial theory by Tumin and Reshotko (2001)~\cite{tumin2001spatial} based on locally parallel flow approximation. The non-parallel effect of boundary layer was incorporated using parabolic stability equations in \cite{zuccher2006parabolic, tempelmann2012spatial, paredes2016optimal, paredes2017instability}, and Paredes et al. (2017)~\cite{paredes2017instability} investigated the nonlinear interaction of the optimally growing streaks and oblique waves. In contrast, the resolvent analysis investigates the properties of the linearized N--S operator for signals with a constant frequency~\cite{trefethen1993hydrodynamic}, and it was extended to global analysis for compressible boundary layers~\cite{bugeat20193d}. Bugeat et al. (2019)~\cite{bugeat20193d} conducted global resolvent analysis on a supersonic boundary layer and identified multiple peaks of amplification in temporal frequency and spanwise wavenumber space, corresponding to the amplification mechanism of streaks, the Mack first mode, and second mode. 

The full nonlinear effect for triggering turbulence is incorporated in stability analysis by nonlinear nonmodal analysis, where Cherubini et al. (2010, 2011)~\cite{cherubini2010rapid, cherubini2011minimal} and Pringle and Kerswell (2010)~\cite{pringle2010using} formulated a variational framework to identify the minimal seed (i.e., the least-energy solution that triggers transition) to depict the bypass transition. For an incompressible boundary layer, Cherubini et al. (2011)~\cite{cherubini2011minimal} investigated the bypass transition path in detail, demonstrating that a self-sustaining process (SSP) involving a disturbance-regeneration cycle operates across multiple scales during transition. However, studies addressing finite-amplitude disturbances in compressible flows remain insufficient compared to incompressible flows. Huang and Hack (2020)~\cite{huang2020variational} extended these approaches to compressible flow fields and demonstrated an increase in disturbance amplification with an increase in the subsonic Mach number in pipe flows. For a supersonic boundary layer at $M=4.5$, Jahanbakhshi and Zaki~\cite{jahanbakhshi2019nonlinearly} applied an ensemble-variational approach of nonlinear non-modal analysis and reported that the least-energy transition path is attributed to the coupling of acoustic modes and oblique modes. Nevertheless, existing studies primarily focus on high Reynolds numbers where multiple transition processes coexist, leaving the nonlinear disturbance amplification mechanisms responsible for early transitions in compressible boundary layers insufficiently explored. 

The objective of the present study is to identify the minimal seed in a supersonic boundary layer at $M=3.0$ and a low Reynolds number, aiming to reveal the early nonlinear mechanism to trigger the transition in supersonic boundary layer. Section \ref{sec:Methodology} details the approaches to obtain stable optimization solution in compressible flows with a sufficient numerical validations. Section \ref{sec:Base flow} presents the properties of base flow fields. The detailed analysis of optimal perturbations and its time-evolution properties are presented in Section \ref{sec:Optimal perturbations and amplification dynamics}.

\section{Methodology}
\label{sec:Methodology}

\subsection{Governing equation}

The governing equations are the non-dimensionalized compressible Navier--Stokes (N--S) equations, expressed as follows:  
\begin{align}
\partial_t \rho + \partial_j (\rho u_j) &= 0, \\ 
\partial_t m_i + \partial_j (m_i u_j) &= - \partial_i p + \partial_j \tau_{ij}, \\
\partial_t p + \partial_j (p u_j) &= - \tilde{\gamma} (p \partial_j u_j - \tau_{ij} \partial_j u_i + \partial_j \Theta_j),
\end{align}
where $\rho, m_i = \rho u_i, u_i, p$ represent the density, $i$-th momentum component, $i$-th velocity component, and pressure, respectively. Here, $\gamma$ is the specific heat ratio, and $\tilde{\gamma} = \gamma-1$. For non-dimensionalization, the standard reference quantities are the freestream density $\rho_e$, velocity $U_e$, temperature $T_e$, and dynamic viscosity $\mu_e$, alongside the Blasius length scale $l$ at a reference location. Given the distance from the leading edge to the reference location, $l_0$, the length scale $l$ and the Reynolds number $Re$ are defined as 
\begin{equation}
l = \sqrt{\frac{\mu_e l_0}{\rho_e U_e}},  \quad Re=\frac{\rho_e U_e l}{\mu_e}. 
\end{equation}

Assuming Stokes' hypothesis, the viscous stress tensor is defined as 
\begin{equation}
\tau_{ij} = \frac{\mu}{Re} \left(\partial_j u_i + \partial_i u_j - \frac{2}{3} \delta_{ij} \partial_k u_k \right), 
\end{equation}
and $\mu$ is the coefficient of viscosity. The heat flux vector is given by  
\begin{equation}
\Theta_i = -\frac{\kappa}{\tilde{\gamma} M^2 Re Pr} \partial_i T, 
\end{equation}
where $M$ is the Mach number, $Pr$ is the Prandtl number, and $\kappa$ is the coefficient of thermal conductivity. To close the system of equations, the equation of state is given in non-dimensional form as 
\begin{equation}
p = \frac{\rho T}{\gamma M^2}
\end{equation}
where $T$ is the temperature. Specifically, we set $Pr=0.7$ and $\gamma=1.4$ assuming a standard atmosphere composition.The coefficient of viscosity is modeled using Sutherland's law. The reference temperature for the model equations is set to the freestream temperature, corresponding to a stagnation temperature of $600 ^\circ R$ ($333 \text{ K}$) following Malik (1990)~\cite{malik1990numerical}. Specific physical parameters, including the reference length $l_0$ and reference temperature $T_e$, are described in detail in the subsequent section on flow conditions. 

\subsection{Nonlinear non-modal analysis}

The methodology of nonlinear non-modal analysis has been well established by Cherubini et al.~\cite{cherubini2010rapid} and Pringle and Kerswell~\cite{pringle2010using}. Our implementation is based on our previous study~\cite{taniguchi2024nonlinear}, with modifications to the selection of the evaluation function and the update procedure for the spatial distribution of the initial disturbance. In this analysis, the total flow field $\bm{q} = \bm{q} (t, \bm{x})$ is decomposed into a steady base flow $\bm{Q} = \bm{Q} (\bm{x})$ and a disturbance flow field $\bm{q}' = \bm{q}' (t, \bm{x})$ as 
\begin{equation}
\bm{q} (t, \bm{x}) = \bm{Q} (\bm{x}) + \bm{q}' (t, \bm{x}). 
\end{equation}
The state vector is defined as $\bm{q} = (\rho, m_x, m_y, m_z, p)^{\top}$. The base flow $\bm{Q}$ is assumed to be a steady solution to the compressible N--S equations. To distinguish these components, the base flow and disturbance fields are denoted as $\bm{Q} = (\bar{\rho}, \bar{m}_x, \bar{m}_y, \bar{m}_z, \bar{p})^{\top}$ and $\bm{q}' = (\rho', m_x', m_y', m_z', p')^{\top}$. The evolution of the perturbed flow is treated as an initial value problem, where the initial state is given by $\bm{q} (0, \bm{x}) = \bm{Q} (\bm{x}) + \bm{\hat{q}}_0$. Here, $\bm{\hat{q}}_0 = \bm{q} ' (0, \bm{x})$ denotes the initial disturbance, which is the subject of optimization in the nonlinear non-modal analysis. 

The Cartesian coordinate system is defined such that $x$, $y$, and $z$ denote the streamwise, wall-normal, and spanwise directions, respectively. The origin in the $xy$-plane is located at the inflow boundary of the computational domain, while the origin of the $z$-coordinate is set arbitrarily. 

Under this configuration, the objective of the nonlinear non-modal analysis is to optimize the initial disturbance field $\bm{\hat{q}}_0$ to achieve maximum amplification at the evaluation time $t=t_f$. To quantify the disturbance magnitude, we introduce an evaluation function representing the disturbance energy for the compressible flow field~\cite{hanifi1996transient} as 
\begin{equation}
\label{equ:def_dis_ene}
E(t) := \frac{1}{2 L_z} \int_{\Omega} \frac{\bar{p}}{\bar{\rho}^2} \rho'^2 + \bar{\rho} u_i' u_i' + \frac{\bar{\rho}}{\gamma \tilde{\gamma} \bar{T} M^2} T'^2 \mathrm{d}V 
\end{equation}
where $\Omega$ denotes the computational domain and $L_z$ is the spanwise length. The amplification of disturbance is defined as the ratio of the disturbance amplitude at the evaluation time to the initial amplitude, as 
\begin{equation}
\label{equ:amp_ratio}
G(t_f) = \frac{E(t_f)}{E(0)}. 
\end{equation}
The optimization maximizes the disturbance amplitude at $t=t_f$ under the constraint of a constant initial disturbance magnitude, $E(0) = \delta^2$, and the condition that the perturbed flow field satisfies the compressible N--S equations as 
\begin{equation}
\partial_t \bm{q} = \mathcal{F} [\bm{q}], 
\end{equation}
where $\mathcal{F}$ symbolically represents the right-hand side of the governing equations. Following the optimization process, the disturbance amplification is evaluated in an extended computational domain with sufficient width to prevent the outflow of disturbances. 

Based on these conditions, the Lagrange multiplier method is applied to optimize disturbance amplification subject to the prescribed constraints. The resulting Lagrangian functional is defined as 
\begin{equation}
\mathcal{L} = E(t_f) - \nu (E(0)-\delta^2) - \int_0^{t_f} \langle \bm{\lambda}, \partial_t \bm{q} - \mathcal{F} [\bm{q}] \rangle \mathrm{d}t
\end{equation}
where $\nu$ and $\bm{\lambda}$ are the Lagrange multipliers, and the brackets $\langle \cdot, \cdot \rangle$ denote the standard $L_2$ inner product. 

The spatial distribution of the optimal perturbation is updated using the steepest-ascent method. By taking the first variation of $\mathcal{L}$ with respect to the state vector $\bm{q}$ and the multipliers, we obtain the Euler--Lagrange equations, which comprise the compressible N--S equations and corresponding adjoint equations. The exact adjoint equation was derived by Ohmichi and Yamada~\cite{ohmichi2021matrix} with adjoint boundary conditions, which are summarized in Appendix \ref{sec:Adjoint compressible N--S equation}. The terminal condition for the adjoint vector at $t=t_f$ is given by 
\begin{equation}
\bm{\lambda} (t_f) = \left( \frac{\partial E(t_f)}{\partial \bm{q}' (t_f)} \right)^{\top}, 
\end{equation}
where the gradient of the energy functional $\partial E/\partial \bm{q}'$ is derived as 
\begin{equation}
\frac{\partial E}{\partial \bm{q}'} = \left(     
(\frac{\gamma}{\tilde{\gamma}} \frac{\bar{p}}{\bar{\rho}^2} + \frac{\bar{u}_i^2}{\bar{\rho}}) \rho' - \frac{\bar{u}_i}{\bar{\rho}} m_i' - \frac{p'}{\tilde{\gamma} \bar{\rho}}, \frac{m_i' - \rho' \bar{u}_i}{\bar{\rho}}, \frac{\gamma M^2 p' - \rho' \bar{T}}{\tilde{\gamma} \bar{\rho} \bar{T}}
\right)^{\top}. 
\end{equation}

The variation of the Lagrangian with respect to the initial disturbance field requires careful treatment in the compressible case. Since the disturbance energy is a quadratic form, we can introduce a weight matrix $A=A(\bm{Q})$ such that $\partial E(0)/\partial \bm{\hat{q}} = A(\bm{Q}) \bm{\hat{q}}$. Especially for low Mach numbers, this matrix $A$ becomes ill-conditioned and it yields the optimization subject to a hyper-elliptical constraint rather than the hyper-sphere in incompressible case, where $A$ simplifies to an identity matrix. To avoid this issue, we introduce a change of variables of $\bm{\hat{q}}_A := \sqrt{A} \bm{\hat{q}} (0)$ to map the disturbance into a space with hyper-spherical constraint. The square root of matrix $A$ can be obtained as follows: 
\begin{equation}
\sqrt{A} = \begin{pmatrix}
   \sqrt{\bar{p}}/\bar{\rho} & 0 & 0 & 0 & 0 \\
   - \bar{u}_x/\sqrt{\bar{\rho}} & 1/\sqrt{\bar{\rho}} & 0 & 0 & 0 \\
   - \bar{u}_y/\sqrt{\bar{\rho}} & 0 & 1/\sqrt{\bar{\rho}} & 0 & 0 \\
   - \bar{u}_z/\sqrt{\bar{\rho}} & 0 & 0 & 1/\sqrt{\bar{\rho}} & 0 \\
   - \sqrt{\bar{p}}/(\bar{\rho} \sqrt{\tilde{\gamma}}) & 0 & 0 & 0 & 1/(\sqrt{\tilde{\gamma} \bar{p}}) \\ 
\end{pmatrix}. 
\end{equation}

The variation with respect to $\bm{\hat{q}}_A$ yields 
\begin{equation}
\label{equ:q0_var}
\left\langle (\sqrt{A}^{-1})^{\top} \bm{\lambda} (0) - \nu \bm{\hat{q}}_A, \delta \bm{\hat{q}}_A \right\rangle = 0. 
\end{equation}
From $E(0)=\delta^2$ and (\ref{equ:q0_var}), the multiplier $\nu$ can be determined as $\nu = \langle (\sqrt{A}^{-1})^{\top} \bm{\lambda} (0), \hat{\bm{q}}_A \rangle/ \| \hat{\bm{q}}_A \|_2^2$ when $(\sqrt{A}^{-1})^{\top} \bm{\lambda} (0)$ is projected onto the space parallel to $\hat{\bm{q}}_A$. Because this gradient of the Lagrangian with respect to $\bm{\hat{q}}_A$ is normal to the constraint plane with a constant energy $E(0)=\delta^2(=\langle \bm{\hat{q}}_A, \bm{\hat{q}}_A\rangle)$, optimization techniques common to the incompressible case \cite{pringle2012minimal, duguet2013minimal} can be applied to update the optimal perturbations. 
\begin{equation}
\label{equ:update_disturbance}
\bm{\hat{q}}_A^{(n+1)} = \bm{\hat{q}}_A^{(n)} + \epsilon^{(n)} \left( (\sqrt{A}^{-1})^{\top} \bm{\lambda}^{(n)} (0) - \nu^{(n)} \bm{\hat{q}}_A^{(n)} \right), 
\end{equation}
where $n$ denotes the optimization step. The optimization process is monitored by computing the residual, $(\sqrt{A}^{-1})^{\top} \bm{\lambda}^{(n)} (0) - \nu^{(n)} \bm{\hat{q}}_A^{(n)}$, which should approach zero upon convergence. To determine the step size $\epsilon$, we applied the momentum-based approach in Pringle and Kerswell~\cite{pringle2012minimal}.

\subsection{Flow configurations}

For the problem setup, we consider a laminar supersonic boundary layer at $M=3.0$. The inflow boundary of the computational domain is set at $x=0$, while the leading edge of the flat plate is located at $x = -300$. The characteristic length for the non-dimensionalization is chosen as $l_0 = 300$ representing the distance between these two points. The computational domain in the wall-normal direction is set to $0 \leq y \leq 100$, which is approximately ten times the typical boundary layer thickness at reference scale for $M=3$. For the nonlinear non-modal analysis, the computational domain is defined as $[x, y, z] \in [0, 450] \times [0,100] \times [-25,25]$. The spanwise extent is chosen to accommodate two periods of the most amplified mode reported by Hanifi et al. (1996)~\cite{hanifi1996transient}. The streamwise size of computational domain for non-modal analysis is carefully selected to retain the disturbance field within the computational domain for target range of evaluation times. 

The numerical scheme is adjusted to obtain a physically reasonable base flow field. The inflow condition of the main calculation is determined via a 2D pre-calculation. This pre-calculation utilizes a self-similar solution under the parallel-flow approximation at $x=-225$ and $Re=150$, obtained through the Illingworth transformation as described in the textbook by White~\cite{white1991viscous}. The numerical implementation of self-similar solution was validated by comparison with the displacement thickness in the results of Malik (1990)~\cite{malik1990numerical}. This pre-computation is conducted for $(x, y) \in [-225, 150] \times [0,100]$ to mitigate inconsistencies between the self-similar solution and the compressible N--S equation, thereby preventing the reflection of numerical shock waves that could alter the boundary layer. For this purpose, a cubic sponge layer \cite{bodony2006analysis} is applied for $75 \leq y \leq 100$, using the uniform flow outside the boundary layer as a reference state. The boundary condition for $y=0$ is an isothermal no-slip condition, while Neumann conditions are applied at the outflows. The calculated one-dimensional profile at $x=0$ is then applied as the inflow boundary condition to time-evolve the flow field. 

In the main calculation domain for the non-modal analysis, the base flow field is obtained from the 2D compressible N--S equations with the inflow boundary condition provided by the pre-calculation, where the sponge region for $75 \leq y \leq 100$ is also applied. The boundary conditions at $y=0$ and $y=100$ are identical to those in the pre-calculation, and a Neumann outflow condition is imposed at $x=450$. To prevent unphysical reflection of disturbances in the 3D non-modal analysis, an additional sponge region based on the 2D base flow is applied at $425 \leq x < 450$, alongside the sponge region for $75 \leq y \leq 100$. This formulation of the sponge region is also incorporated into the adjoint equations in Appendix \ref{sec:Adjoint compressible N--S equation}. Periodic boundary conditions are imposed in the spanwise direction at $z = \pm 25$. 

The time evolution of the induced disturbance field is traced in the computational domain, which is extended into the streamwise direction $x \in [0, 1350]$. The amplification of the disturbance in (\ref{equ:amp_ratio}) is evaluated in this extended computational domain.

\subsection{Numerical discretization methods}

In the present study, we apply consistent numerical schemes for both compressible N--S equations and their adjoint equations. Spatial derivatives are evaluated using a sixth-order compact finite difference scheme~\cite{lele1992compact}. To suppress numerical aliasing instabilities arising from poorly resolved high-frequency modes, a tenth-order compact filtering method~\cite{gaitonde2000pade} is applied ($\alpha_f = 0.45$ for the filtering strength). Time integration is performed using a third-order Total Variation Diminishing (TVD) Runge--Kutta scheme~\cite{gottlieb1998total}. In the time-evolution of adjoint equations in non-modal analysis, the checkpoints method was applied to preserve and recalculate all the flow fields of each time step, except for the linear interpolation of the intermediate stage inside the Runge-Kutta method. 

The computational domain for the nonlinear non-modal analysis is discretized with a grid of $640 \times 256 \times 128$ points in the streamwise ($x$), wall-normal ($y$), and spanwise ($z$) directions, respectively. For the extended domain, the spanwise grid is retained and smoothly connected to the computational domain for non-modal analysis, where the total grid number is $1920 \times 256 \times 128$ points. Uniform grid spacing is employed in the spanwise direction, with $\Delta z=0.391$. In the wall-normal direction, the grid is clustered near the wall surface; the minimum spacing is $\Delta y_{\mathrm{min}} = 0.0648$, which corresponds to approximately $1.2 \%$ of the displacement thickness $\delta^*$ under adiabatic wall conditions. Based on the turbulent skin friction by Van Driest II~\cite{van1956turbulent, white1972simple}, this corresponds to $y^+ = 0.36$ at the outer edge of the computational domain for non-modal analysis ($x=450, Re=474$) and $y^+ = 0.49$ at the downstream edge of the extended domain ($x = 1350, Re = 704$), respectively. This spatial resolution is verified to be sufficient for resolving the characteristic wavelengths of Mack's first and second modes, as well as the resulting nonlinear interactions.

\subsection{Numerical validation}

To validate the numerical framework, we computed nonlinear optimal perturbations (NLOPs) in the incompressible limit and compared them with the results by Cherubini et al. (2011)~\cite{cherubini2011minimal}. To align with the problem setup described in the literature, we set the computational domain to $[0,200] \times [0,20] \times [0,10.5]$. Coordinates were normalized by the boundary-layer thickness $d$ at the corresponding Mach number. The computational domain was discretized using a grid of $768 \times 192 \times 96$ points with a minimum wall spacing of $\Delta y = 0.1$, consistent with their approach. We set the Reynolds number based on the boundary layer thickness, $Re_{d}=300$, for a Mach number of $M=0.3$. 

\begin{figure}[ht]
\centering
\includegraphics[width=7cm]{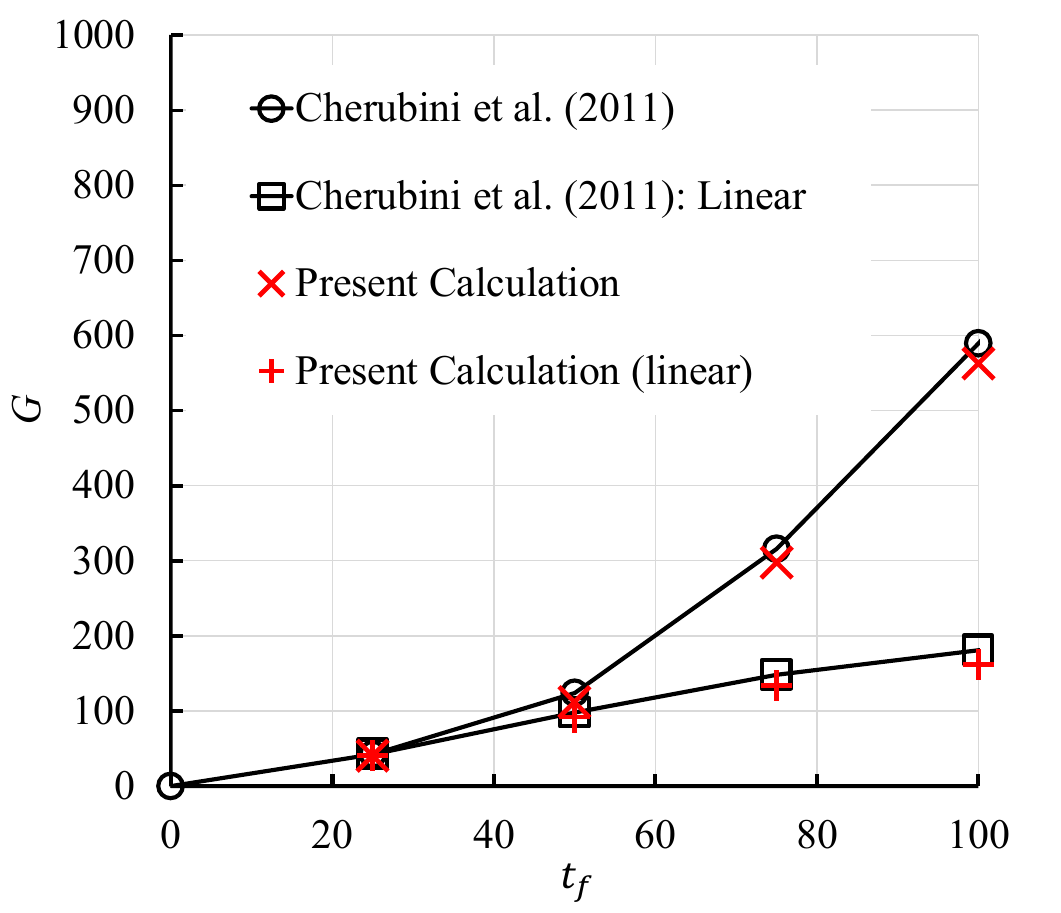} 
\caption{\label{fig:pic_incomp_cher}
Comparison of growth rates of incompressible boundary layer from Cherubini et al. (2011)~\cite{cherubini2011minimal} and  results for a subsonic Mach number at $M=0.3$ in present study. 
}
\end{figure}

Cherubini et al. (2011)~\cite{cherubini2011minimal} used the kinetic energy norm $E_{\mathrm{incomp.}} = \int_{\Omega} u_i' u_i' \mathrm{d}V$ for incompressible flow fields. The initial disturbance amplitude is set to $\delta = 10^{-3}$ for the linear limit and $\delta = 0.05$ for the nonlinear case, which corresponds to $\delta = 0.1$ in their work due to a constant difference between setups. Given that our study includes the effect of density and pressure fluctuations on disturbance amplitudes, we note that the current setting even at low subsonic Mach numbers is not exactly the same as the incompressible case. Figure \ref{fig:pic_incomp_cher} compares the present results with those in the literature, demonstrating excellent agreement. For longer evaluation times $(t_f = 75, 100)$, both linear and nonlinear optimal perturbations showed slight deviations from the literature values, resulting in an underestimation of $4.6 \%$ at $t_f = 100, \delta = 0.05$. This validates the present optimization scheme with respect to the incompressible limit. 

Furthermore, we consider the effect of the outflow boundary conditions on the disturbance amplification. Table \ref{tab:dif_growth} shows the difference in the amplification rate of the identified NLOP between the original and extended computational domains. This table demonstrates that no disturbance reflection occurs at the outflow boundary, which could otherwise cause spurious amplification. 

\begin{table}[ht]
\caption{\label{tab:dif_growth}
Comparison of growth rates for the NLOPs in the computational domain of $x \in [0,450]$ and $x \in [0,1350]$ at $t_f = 200$ for adiabatic wall conditions. 
}
\begin{tabular}{c|ccc}
$\delta$ & $G_{[0,450]}$ & $G_{[0,1350]}$ & Relative error \\ \hline
$0.01$ & $206.91$ & $206.91$ & $-3.8\times 10^{-3} \%$ \\ 
$0.02$ & $233.12$ & $233.09$ & $1.4 \times 10^{-2} \%$ \\ 
$0.05$ & $336.58$ & $336.41$ & $5.0 \times 10^{-2} \%$
\end{tabular}
\end{table}

To investigate the grid convergence of the present analysis, we compare the current numerical grid of $640 \times 256 \times 128$ with a coarse grid of $480 \times 192 \times 96$ points and a fine grid of $800 \times 320 \times 160$ points, while keeping the grid stretching ratios and spatial distribution identical. Table \ref{tab:grid_convergence} shows a comparison of disturbance amplification at $t_f = 200$ and $T_w = 2.5$ with different initial disturbance amplitudes, where the current grid converges within a relative error of $2 \%$ compared with the coarse and fine grids except for the case at $\delta = 0.04$. Although the optimization process has a high sensitivity to the adjoint equations, the resolution of the numerical grid does not significantly affect the disturbance growth, except for the intermediate disturbance amplitude cases. Appendix \ref{sec:Grid dependency of NLOP} further investigates the cause of this discrepancy especially at $\delta = 0.04$, and this discrepancy is attributed to being trapped in a local optimum rather than the grid dependency of the numerical results. Indeed, in this case of $\delta = 0.04$, the relative error between the coarse and fine grids remains at $0.18 \%$, indicating the grid convergence when settling into a single optimum. A further increase in initial disturbance amplitude shows excellent agreement despite the difference in the number of grid points. 

\begin{table}[ht]
\caption{\label{tab:grid_convergence}
Dependency of numerical grid resolution of disturbance amplification at $t_f = 200, T_w = 2.5$. $R_{\mathrm{coarse}}, R_{\mathrm{fine}}$ indicates the relative error of each coarse and fine case from the current grid, respectively. 
}
\begin{tabular}{c|c|cc|cc} 
$\delta$ & $G_{\mathrm{current}}$ & $G_{\mathrm{coarse}}$ & $R_{\mathrm{coarse}}$ & $G_{\mathrm{fine}}$ & $R_{\mathrm{fine}}$ \\ 
\hline
$2.0\times10^{-5}$ & $205.6$ & $205.3$ & $-0.10 \%$ & $203.4$ & $-1.07 \%$ \\
$0.02$ & $233.1$ & $237.2$ & $1.75 \%$ & $237.5$ & $1.87 \%$\\
$0.04$ & $302.8$ & $285.0$ & $-5.89 \%$ & $285.4$ & $-5.75 \%$ \\
$0.05$ & $336.4$ & $338.7$ & $0.69 \%$ & $337.5$ & $0.34 \%$ \\
\end{tabular}
\end{table}

\section{Base flow}
\label{sec:Base flow}

\begin{figure}[ht]
\centering
\includegraphics[width=6cm]{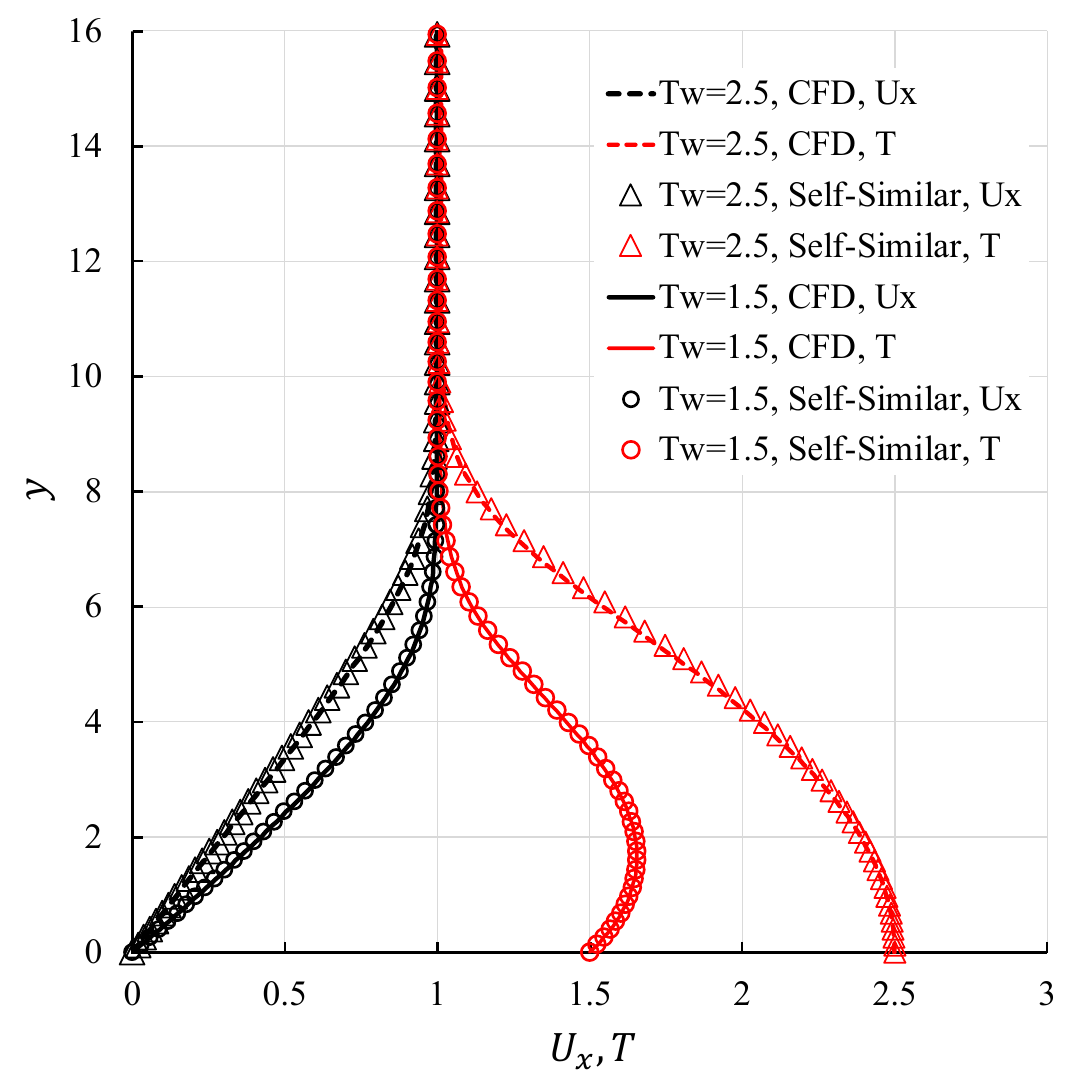} 
\caption{\label{fig:pic_inflow_cond}
Comparison of wall-normal profiles at the inflow boundary for $T_w = T_{ad}, 0.6 T_{ad}$: numerical results (open symbols) versus self-similar solution (lines). 
}
\end{figure}

Figure \ref{fig:pic_inflow_cond} compares the wall-normal profiles of velocity and temperature at the inflow boundary with the self-similar solution. This figure demonstrates excellent agreement between the computed base flow and the self-similar solution, indicating the ability of the numerical framework to absorb numerical inconsistencies from the pre-calculation domain. However, a numerical shock wave generated in the pre-calculation domain causes the pressure at the inflow boundary to slightly exceed the uniform flow pressure of $p_0 = 1/(\gamma M^2) = 0.79$. Specifically, the pressure increases by approximately $4 \%$ at the boundary and gradually decreases in the streamwise direction. These discrepancies are considered to be inevitable under the current computational configuration and selected numerical scheme. Nevertheless, no local pressure fluctuations are observed that would suggest the reflection of a numerical shock wave to impact the base-flow velocity profiles. 

Wall cooling reduces the temperature within the boundary layer, shifting the location of the peak temperature further from the wall. This temperature decrease reduces the local viscosity via Sutherland's law, yielding a thinner boundary layer, which is also well reproduced in the same figure. The fundamental properties of the displacement thickness and the location of the generalized inflection point (GIP) were investigated using the self-similar solution of the compressible boundary layer equations by decreasing the wall temperature in steps of $10 \%$ from the adiabatic condition. Specifically, the displacement thickness decreases by $32.2\%$, from $5.40$ at $T_w = T_{\mathrm{ad.}}$ to $3.66$ at $T_w = 0.6 T_{\mathrm{ad.}}$. Under the adiabatic wall condition, the GIP is located at $y=5.17$. As the wall temperature decreases to $T_w = 0.9 T_{\mathrm{ad.}}$ and $0.8 T_{\mathrm{ad.}}$, the GIP undergoes a bifurcation, and subsequently disappears entirely at $T_w = 0.7 T_{\mathrm{ad.}}$ and $0.6 T_{\mathrm{ad.}}$. Thus, reducing the wall temperature to below $T_w = 0.6 T_{\mathrm{ad.}}$ is sufficient to isolate the specific influences of wall cooling. Linear stability characteristics of base flow is summarized to support the difference of linear stability properties in Appendix \ref{sec:Linear stability properties of base flow}. 

\section{Optimal perturbations and amplification dynamics}
\label{sec:Optimal perturbations and amplification dynamics}

\subsection{Optimization properties}

\begin{figure}[ht]
\centering
\begin{minipage}[b]{0.95\hsize}
\centering
\includegraphics[width=7.0cm]{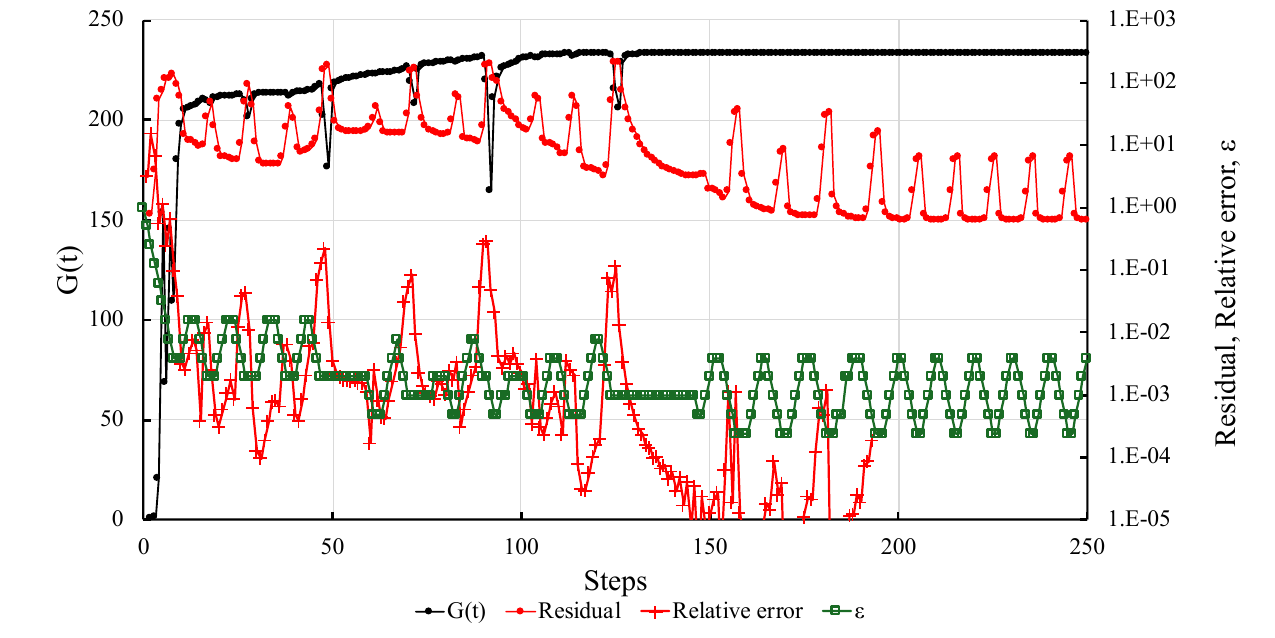}
\subcaption{$\delta = 0.02$} 
\end{minipage} \\
\begin{minipage}[b]{0.95\hsize}
\centering
\includegraphics[width=7.0cm]{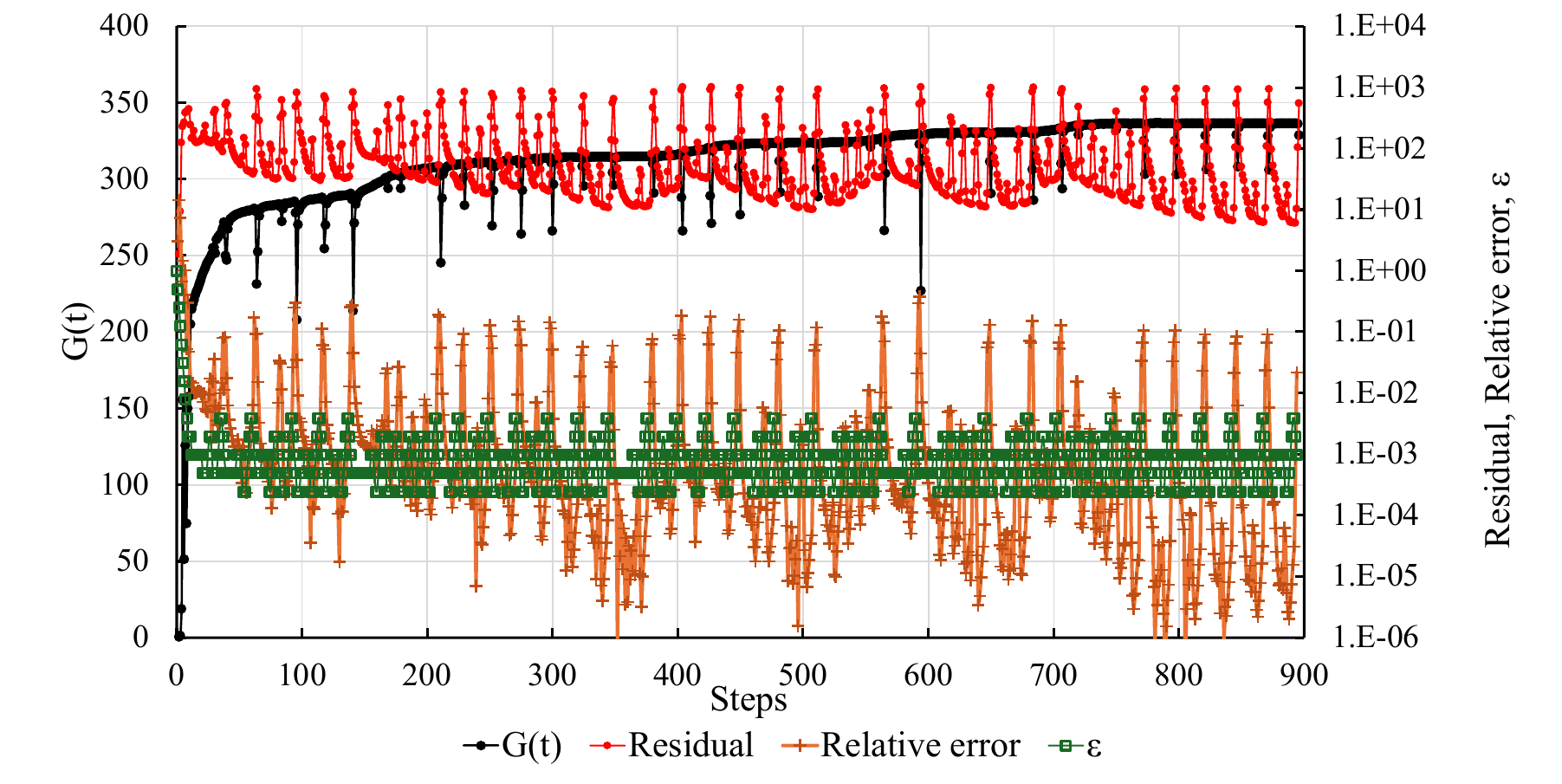}
\subcaption{$\delta = 0.05$}
\end{minipage} 
\caption{\label{fig:pic_convergence}
History of nonlinear non-modal analysis under the adiabatic wall condition with $t_f = 200$. Symbols used are black circles for disturbance amplification, red circles for the residual as the amplitude of the gradient, orange crosses for the relative error of the evaluation function during the optimization and green open squares for the step size $\epsilon$. 
} 
\end{figure}

We first explore the convergence properties of nonlinear non-modal analysis. Figure \ref{fig:pic_convergence} illustrates a typical convergence process in calculating optimal perturbation using the steepest ascent method combined with the continuous adjoint method. During optimization, the residual of the evaluation function in (\ref{equ:update_disturbance}) gradually decreases, leading to an increase in the evaluation function after sufficient optimization steps. A higher initial disturbance amplitude necessitates a larger number of optimization steps for solution convergence. This is evident from comparing Figs. \ref{fig:pic_convergence} (a) and (b), where a downward envelope of the residual forms a plateau of $O(10)$. This limitation of the steepest ascent method has already been pointed out by Olvera and Kerswell (2017)~\cite{olvera2017optimizing}, especially in contexts utilizing transient growth to identify coherent structures in incompressible flows. The absence of self-adjoint properties in numerical discretization and the use of compact filtering methods to eliminate high wavenumber components for solving adjoint equations further compound these limitations. Moreover, Fig. \ref{fig:pic_convergence} exhibits periodic spikes with an overall decreasing trend, which occur after increasing the step size $\epsilon$. This spike behavior originates from steep gradients of the evaluation function in compressible flow fields. 

The convergence criterion for optimization is defined as the relative error falling below $10^{-5}$ across optimization steps. However, the nonlinearity inherent in disturbance growth hinders rapid convergence during the NLOP calculations. This single criterion can lead to premature convergence toward a local optimum, as demonstrated in Fig. \ref{fig:pic_convergence} (b), due to premature stagnation on a local gradient plateau. Since our optimization scheme does not ensure monotonicity, periodic spikes persist throughout the optimization process. To prevent premature termination, optimization is halted when a gradient plateau appears following three distinct spike peaks. 

\subsection{Optimal perturbations}
\label{sec:Optimal perturbations}

\begin{figure}[ht]
\centering
\begin{minipage}[b]{0.45\hsize}
\centering
\includegraphics[width=5cm]{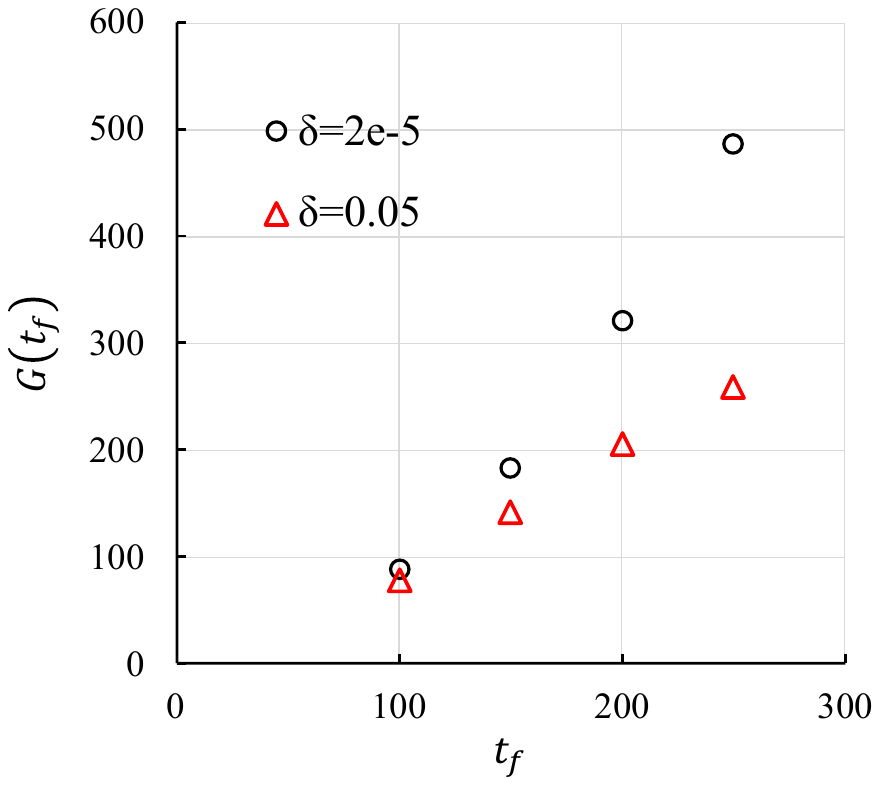}
\subcaption{$T_w = 2.5$}
\end{minipage} 
\begin{minipage}[b]{0.45\hsize}
\centering
\includegraphics[width=5cm]{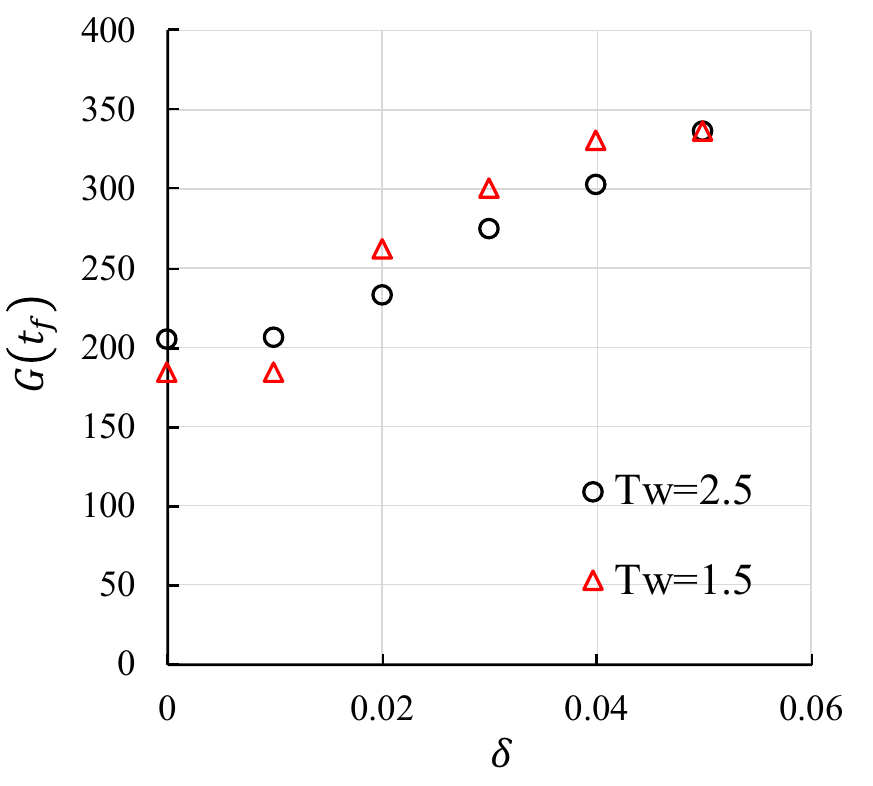}
\subcaption{$t_f = 200$}
\end{minipage} 
\caption{
Growth rate of disturbances at (a) adiabatic wall and with two different initial disturbance amplitudes and (b) the effect of initial disturbance amplitude under adiabatic and cooled temperature conditions at $t_f = 200$. 
}
\label{Fig:pic_amp_time}
\end{figure}

Figure \ref{Fig:pic_amp_time} (a) compares the energy amplifications of linear ($\delta = 2.0\times 10^{-5}$) and nonlinear ($\delta = 0.05$) optimal perturbations under the adiabatic wall condition. For $t_f \geq 100$, nonlinearity enhances the growth rate of the disturbance field due to the nonlinear self-interactions. Figure \ref{Fig:pic_amp_time} (b) illustrates this transition from linear to nonlinear amplification regimes, where a large initial disturbance ($\delta > 0.01$) gradually increases the growth rate from the baseline linear amplification starting from $\delta=0.01$, until numerical divergence occurs at $\delta = 0.06$. Under the cooled wall condition, the linear amplification rate decreases compared to the adiabatic condition, corresponding to the suppression of the Mack first mode~\cite{mack1984boundary} in linear stability theory. In contrast, cooled boundary wall conditions exhibit significant nonlinear enhancement of disturbance growth from a lower initial amplitude than that observed under the adiabatic condition. 

\begin{figure}[ht]
\centering
\includegraphics[width=8cm]{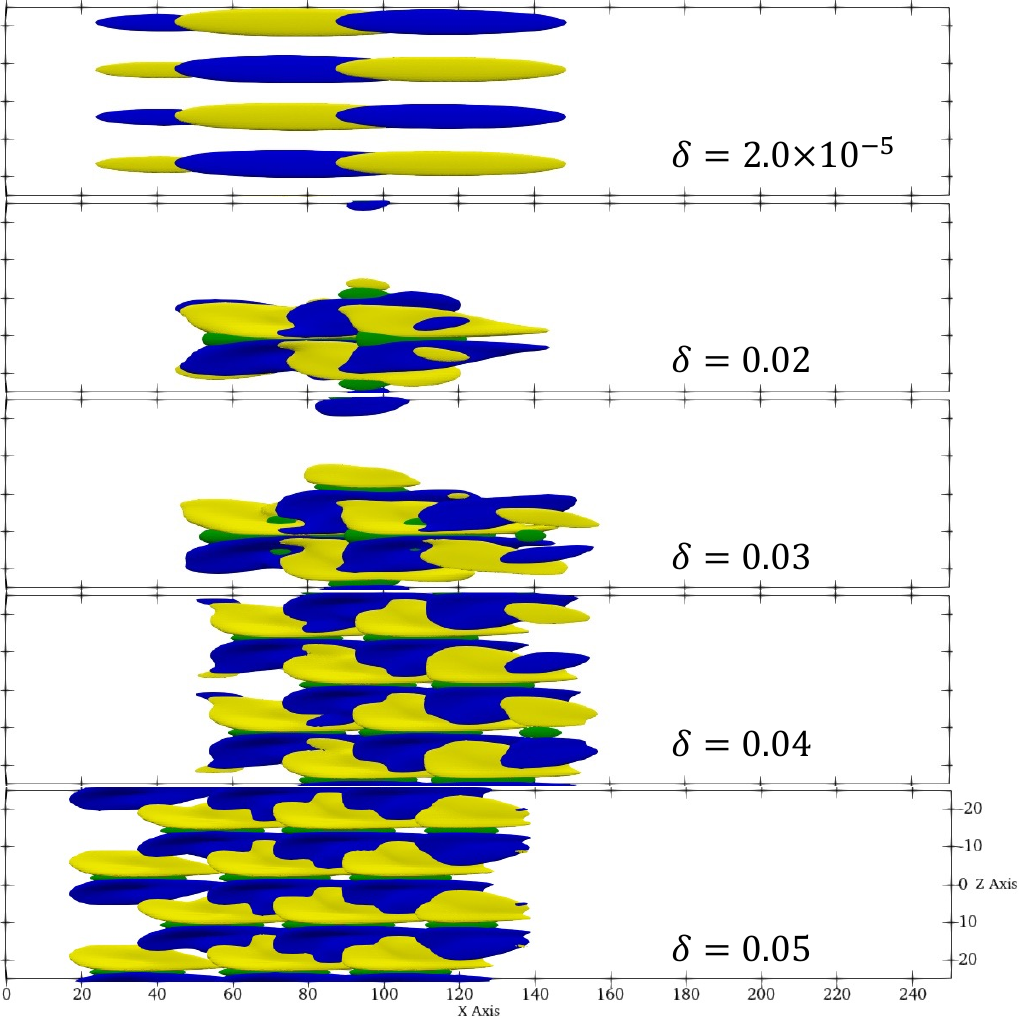} 
\caption{\label{Fig:pic_omega_pent}
Iso-contour of optimal perturbation normalized with its initial disturbance amplitude: $\omega_x' = 0.01$ for yellow (appearing as light grey in grayscale), $\omega_x' = -0.01$ for blue (thick grey), and $u_x = 0.01$ for green (mid-grey). }
\end{figure}

Figure \ref{Fig:pic_omega_pent} illustrates the spatial distribution of the streamwise disturbance vorticity $\omega_x'$ and the streamwise disturbance velocity $u_x'$ for optimal perturbations with varying initial disturbance amplitudes. The dominant frequency is analyzed using discrete Fourier analysis interpolated onto a uniform grid and windowed using a Hann function on the $x$-$z$ distribution of $u_x'$ at $y=4.2, 6.6, 9.57$. Due to the non-parallel effects of the boundary layer, the streamwise and spanwise wavenumbers are spatially averaged over the streamwise domain. 

In the linear case ($\delta = 2\times 10^{-5}$), the streamwise vortex structures exhibit a periodic distribution with alternating positive and negative values in the streamwise direction, consistent with the findings of Cherubini et al. (2011) \cite{cherubini2011minimal} for an incompressible boundary layer. This structure features a dominant wavelength of $(\lambda_x, \lambda_z) = (90.3, 25.0)$, which is also observed at $\delta = 0.01$. Compared to linear stability theory results shown in Fig. \ref{Fig:pic_LST_eig01}, this peak wavenumber lies within the unstable range in the streamwise direction, but exhibits a value roughly twice as large in the spanwise direction. The predominance of the streamwise vorticity also aligns with the global resolvent analysis results by Bugeat et al. (2019)~\cite{bugeat20193d}, who investigated the characteristics of staggered forcing near the parameters of the Mack first mode and streaks in a compressible boundary layer at $M=4.5$. The emergence of this intermediate structure between the Mack first mode and the streak is reasonable in the present study, given the significantly shorter evaluation time ($t_f = 200$) relative to the optimal time ($t_{opt} = 1120$) predicted for the LOP by Hanifi et al. (1996)~\cite{hanifi1996transient}. Consequently, the present NLOP calculated with a small initial disturbance amplitude yields a physically reasonable structure reflecting the non-normal transitional behavior between the Mack first mode and the streak. 

\begin{figure}[ht]
\centering
\begin{minipage}[b]{0.3\hsize} 
\includegraphics[width=3.5cm]{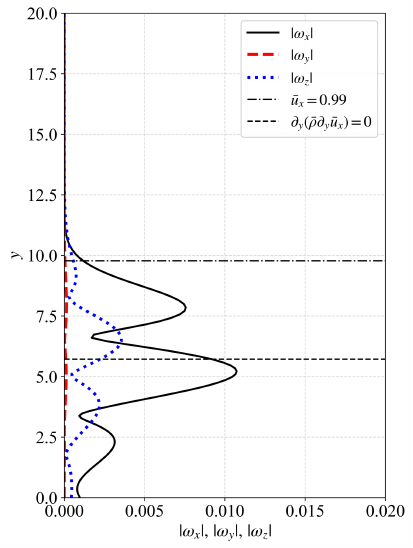}
\subcaption{$\delta =2\times10^{-5}, x=90$}
\end{minipage}
\begin{minipage}[b]{0.3\hsize} 
\includegraphics[width=3.5cm]{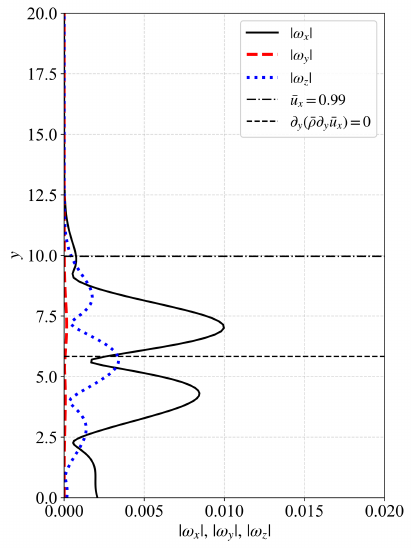}
\subcaption{$x=105$}
\end{minipage}
\begin{minipage}[b]{0.3\hsize} 
\includegraphics[width=3.5cm]{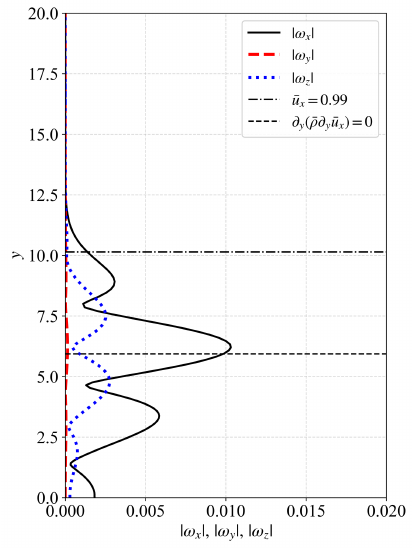}
\subcaption{$x=120$}
\end{minipage}  \\
\begin{minipage}[b]{0.3\hsize} 
\includegraphics[width=3.5cm]{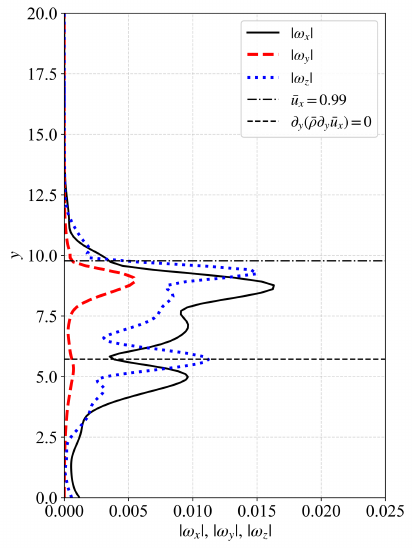}
\subcaption{$\delta =0.04, x=90$}
\end{minipage}
\begin{minipage}[b]{0.3\hsize} 
\includegraphics[width=3.5cm]{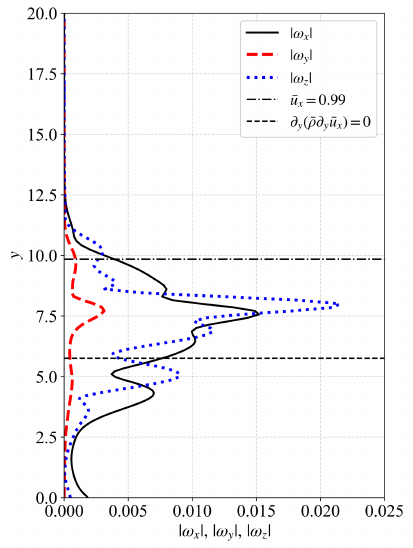}
\subcaption{$x=95$}
\end{minipage}
\begin{minipage}[b]{0.3\hsize} 
\includegraphics[width=3.5cm]{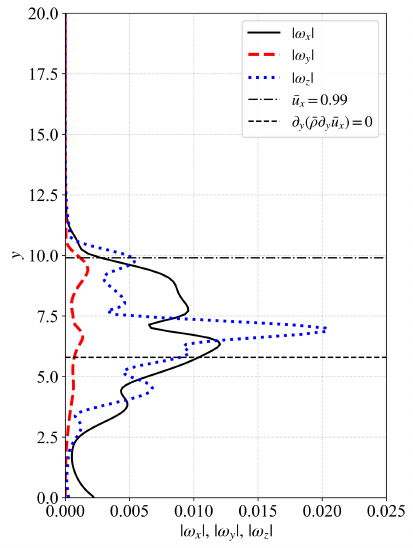}
\subcaption{$x=100$}
\end{minipage}  
\caption{
Wall-normal distribution of RMS for disturbance vorticity in spanwise direction, normalized with its initial disturbance amplitude. (a-c) for linear case $\delta = 2 \times 10^{-5}$ and (d-f) for nonlinear case $\delta = 0.04$. 
}
\label{Fig:pic_lin_90}
\end{figure}

Figure \ref{Fig:pic_lin_90} shows the wall-normal distribution of the spanwise root-mean-square (RMS) of the disturbance vorticity, which characterizes one spanwise period of the optimal perturbation. The linear case demonstrates that the streamwise vorticity dominates the structure of the optimal perturbation and is distributed over a wide range of the boundary layer height. 

An increase in the initial disturbance amplitude leads to overall changes in the spatial distribution due to nonlinear effects, as demonstrated in Fig. \ref{Fig:pic_omega_pent}. The optimal perturbation shows a streamwise-localized distribution with a decrease in its spanwise period. In contrast to the linear case, the dominant streamwise wavelength is significantly shortened and remains nearly constant within a narrow range from $\lambda_x = 36.1$ at $\delta = 0.02$ to $33.5$ at $\delta = 0.05$. The reduction in spatial wavelength from linear prediction is also observed in DNS calculations of low subsonic boundary layers~\cite{yakeno2026new}. The spatial distribution of the NLOP exhibits a structure consisting of two lines of alternating streamwise vorticity closely spaced in the spanwise direction, which deviates significantly from the behavior reported for incompressible flows~\cite{cherubini2011minimal}, where nonlinear structures consist of a single linear optimal perturbation structure shrinking in the streamwise direction. The streamwise velocity component is induced between these two lines of streamwise vorticity and is associated with core streamwise disturbance vorticity, which manifests as the spanwise vorticity $\omega_z'$, as shown in Fig. \ref{Fig:pic_lin_90} (d,e,f). The upward shift in the wall-normal distribution of the optimal perturbation indicates active self-interactions of disturbance vorticity in finite-amplitude optimal perturbations. 

\begin{figure}[ht]
\centering
\includegraphics[width=9cm]{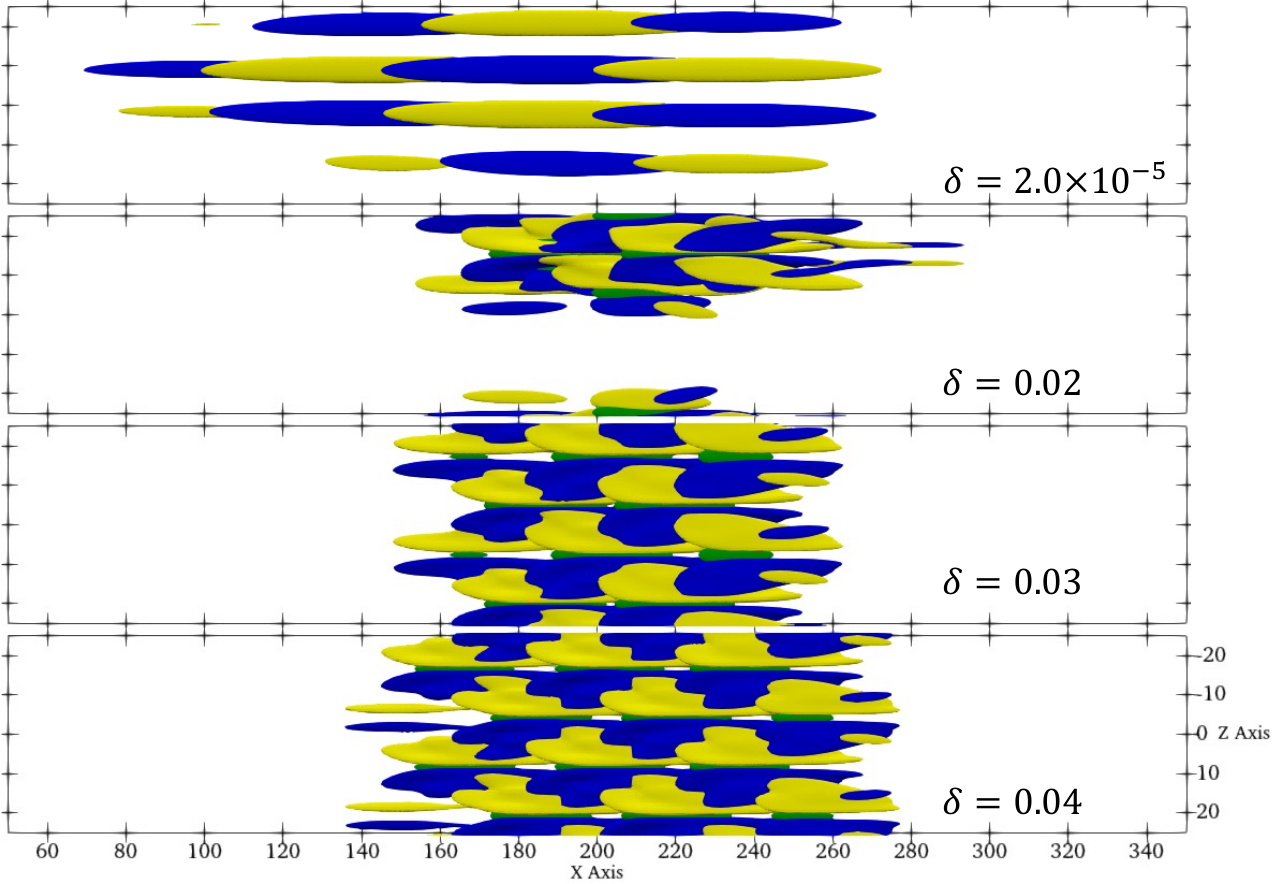} 
\caption{\label{Fig:pic_omega_pent_Tw150}
Iso-contour of optimal perturbation normalized with its initial disturbance amplitude at $T_w = 0.6 T_{\mathrm{ad.}}$. See caption of Fig. \ref{Fig:pic_omega_pent}. }
\end{figure}

Figure \ref{Fig:pic_omega_pent_Tw150} shows the spatial distribution of optimal perturbation under cooled-wall conditions. Compared to adiabatic wall conditions, the spatial distribution along the streamwise direction shifts downstream due to a decrease in boundary layer thickness. The linear case ($\delta = 2.0\times 10^{-5}$) displays similar characteristics to those seen with the adiabatic wall condition but exhibits slightly longer dominant streamwise wavelengths $\lambda_x = 100.4$, even though the generalized inflection point moves outside the boundary layer, as discussed in Fig. \ref{fig:pic_inflow_cond}. The spatial distribution characteristics resemble those under adiabatic wall conditions with a dominant streamwise wavelength ranging from $\lambda_x = 33.5$ at $\delta = 0.02$ to $36.1$ at $\delta = 0.04$, except for the decrease in the threshold value required to trigger changes in spatial distribution, supported by the shifts in the amplification ratio shown in Fig. \ref{Fig:pic_amp_time}. These results suggest that a decrease in wall temperature may enhance nonlinear amplification of disturbances, unlike in linear optimal perturbations. 

\subsection{Time-evolution of disturbances}

\begin{figure}[ht]
\centering
\includegraphics[width=9cm]{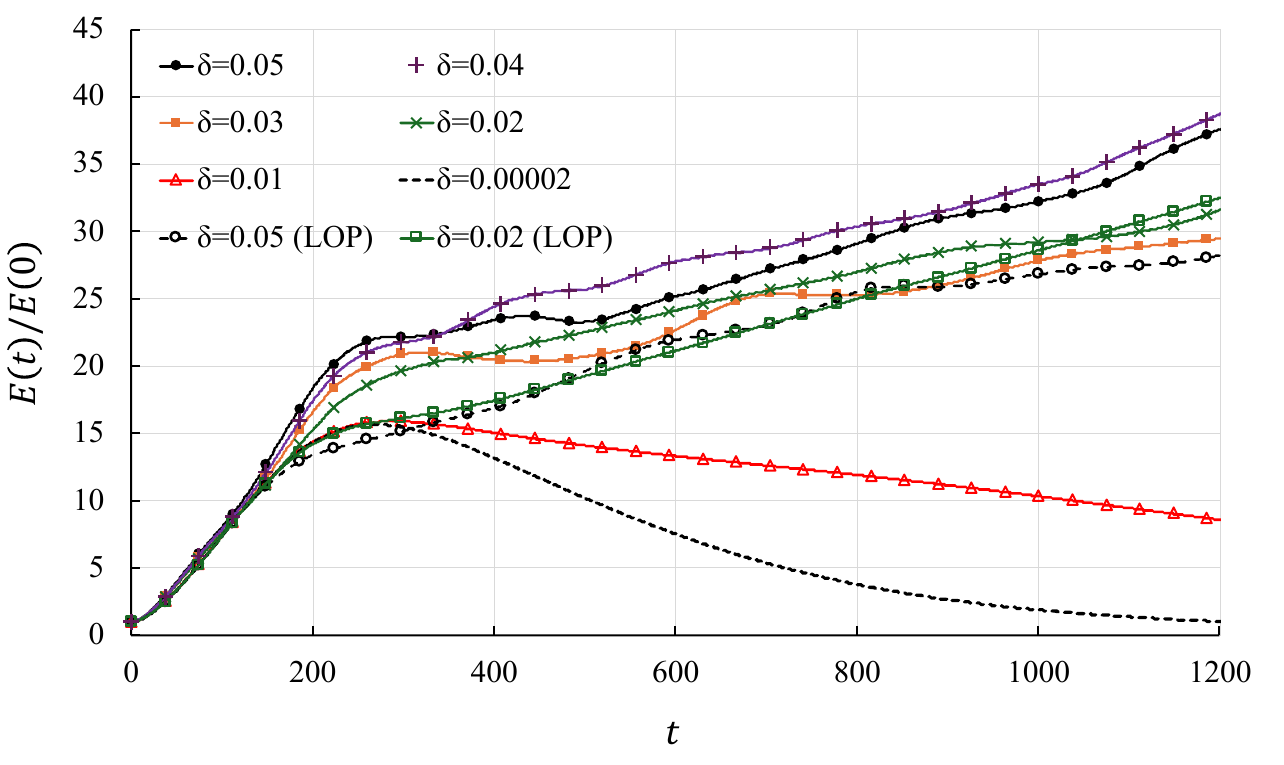} 
\caption{\label{fig:pic_dis_evol}
Time evolution of disturbance field for optimal perturbations with different initial disturbance amplitudes under adiabatic wall temperature condition with $t_f = 200$ for all the conditions. The legend with (LOP) indicates the disturbance calculated at $\delta = 2 \times 10^{-5}$ is scaled up by a constant factor. 
}
\end{figure}

Figure \ref{fig:pic_dis_evol} shows the evolution of the disturbance amplitudes, normalized by their initial magnitude. The linear case of $\delta=2\times 10^{-5}$ exhibits transient amplification. With an increase in initial disturbance amplitude, the initial effect of nonlinearity serves to sustain the highly amplified finite disturbance field after the linear transient growth ceases, as shown in the case with $\delta = 0.01$. A further increase in the initial disturbance amplitude leads to a continuous growth of the amplitude as time progresses. To quantify the nonlinear effects, the same figure also compares the evolution of the linear condition ($\delta = 2 \times 10^{-5}$) scaled up by a constant factor to $\delta = 0.02, \delta = 0.05$, which shows the evolution of disturbances for a longer evaluation time. For the case of $\delta = 0.05$, the growth rate at the evaluation time $t_f=200$ is limited to $179.0$, representing a $13.0 \%$ reduction compared to the linear growth rate, and this value is lower by $46.8 \%$ compared to the NLOP. Furthermore, these scaled-up LOPs delay the increase of their amplitudes compared to the NLOP at the same evaluation time. This point is consistent with the fact that the NLOP selects a different amplification path driven by nonlinear mechanisms distinct from the linear amplification mechanism, as suggested by Kerswell (2018)~\cite{kerswell2018nonlinear}. 

\begin{figure}[ht]
\centering
\begin{minipage}[b]{0.45\hsize}
\centering
\includegraphics[width=5.5cm]{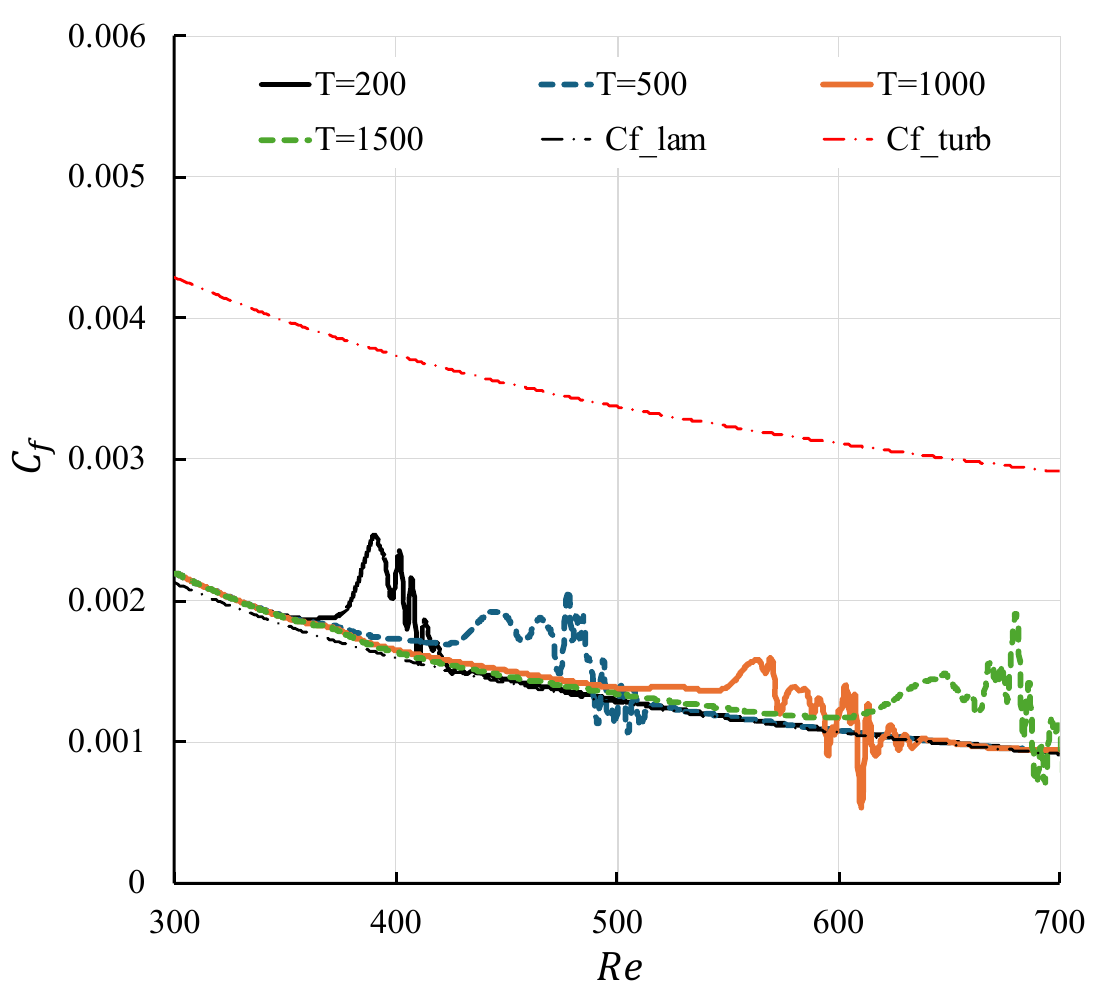}
\subcaption{$\delta = 0.03$}
\end{minipage} 
\begin{minipage}[b]{0.45\hsize}
\centering
\includegraphics[width=5.5cm]{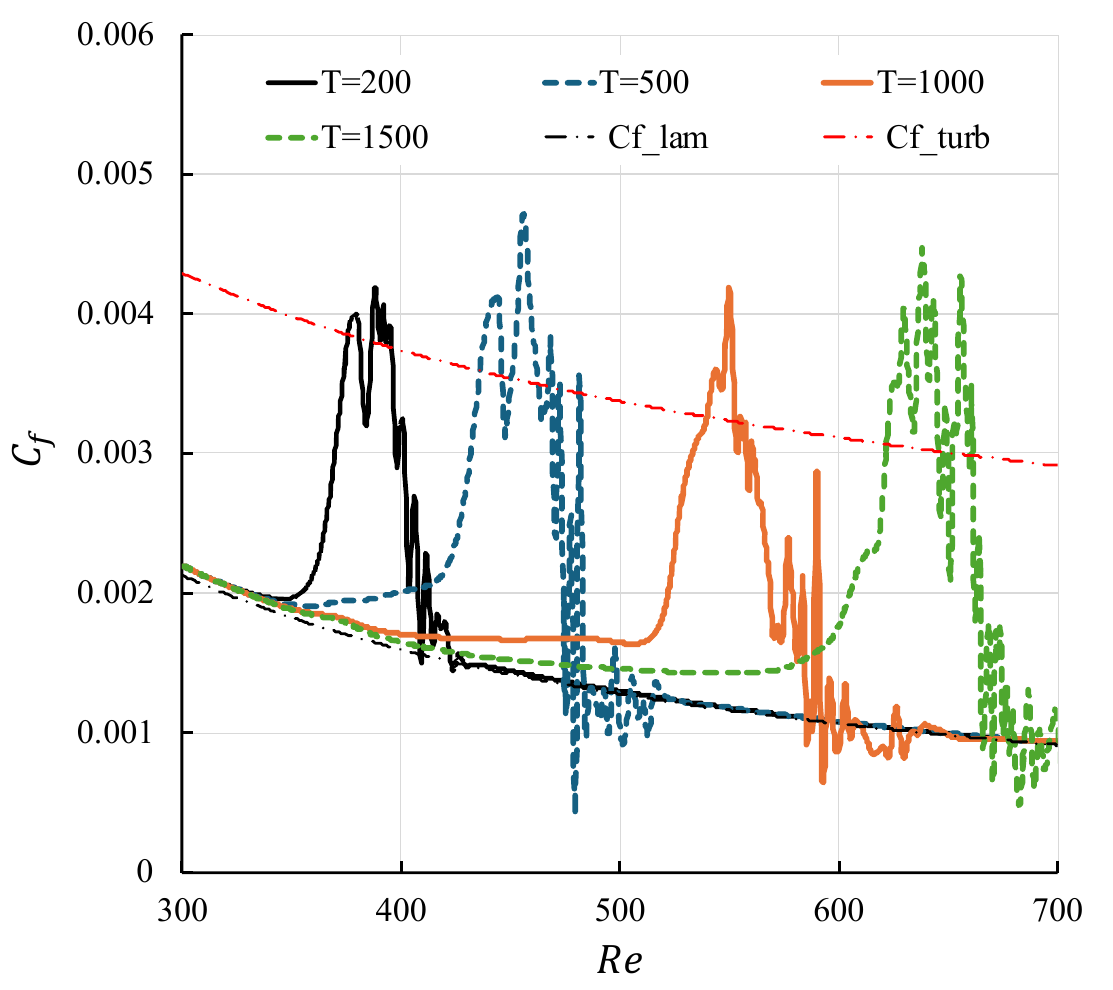}
\subcaption{$\delta = 0.05$}
\end{minipage} 
\caption{
Coefficient of skin friction for instantaneous flow fields at different times under adiabatic wall condition compared with the laminar skin friction $C_{f\mathrm{lam.}}$ and the empirical estimation of Van Driest II $C_{f\mathrm{turb.}}$. 
}
\label{Fig:pic_cf_d150}
\end{figure}

Figure \ref{Fig:pic_cf_d150} shows the evolution of skin friction after the application of NLOPs to the flow field. In this figure, we compare the obtained skin friction with the self-similar solution of compressible boundary layer equations and the empirical turbulence skin friction relations by Van Driest II~\cite{van1956turbulent, white1972simple}. Note that the current low Reynolds number setting in a supersonic boundary layer ($M=3.0, Re=300$) is beyond the standard applicable range of empirical relations assuming the natural transition of the boundary layer. In both cases, a broad increase in skin friction as the disturbances advect downstream is obtained in the present calculation, indicating that homogeneous momentum transfer results in a flattened skin friction profile. However, the degree of increase in skin friction strongly depends on the initial disturbance amplitude, demonstrating the instability of forcibly transitioned turbulent flows. 

\begin{figure}[ht]
\centering
\begin{minipage}[b]{0.95\hsize}
\centering
\includegraphics[width=8cm]{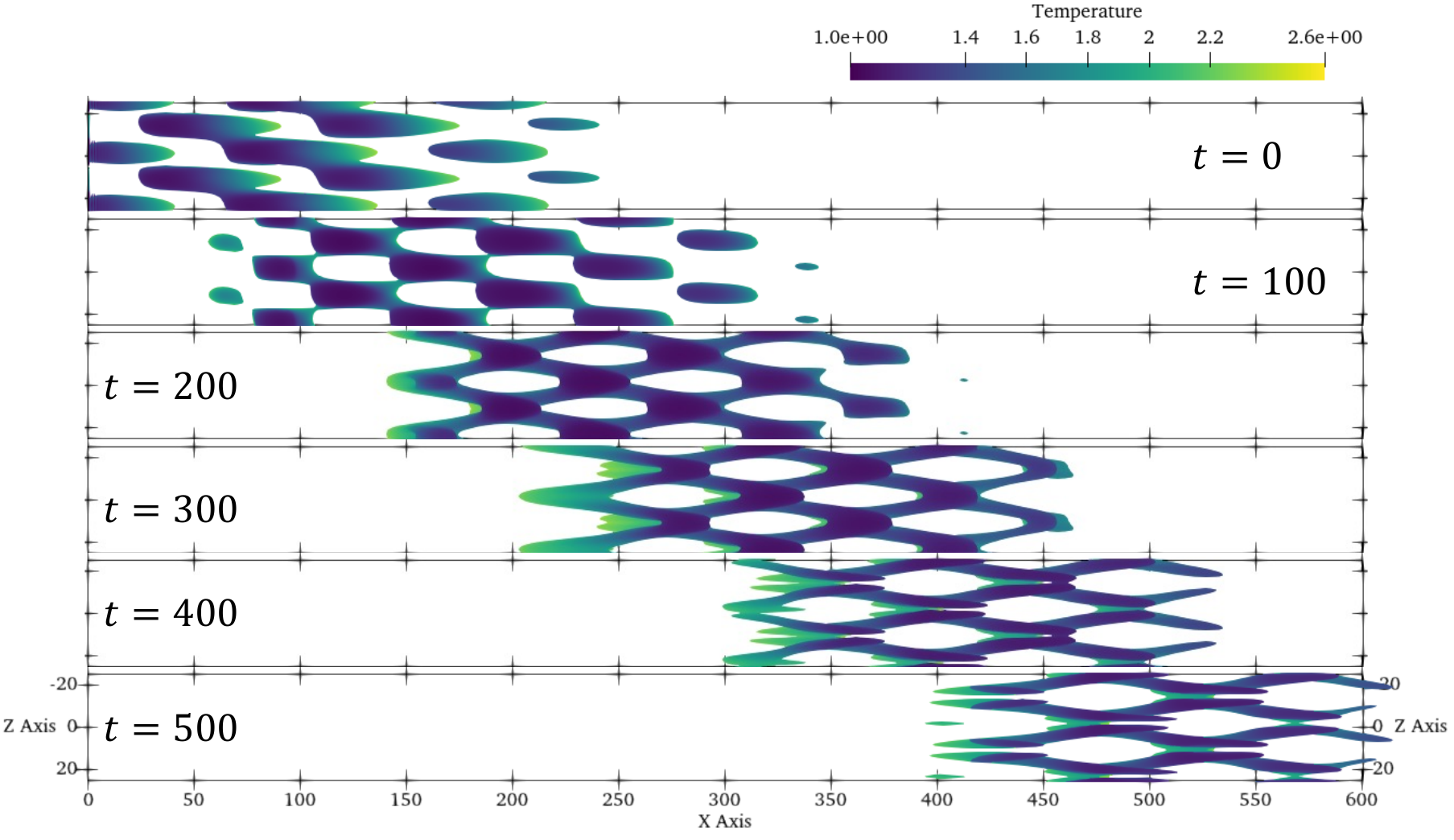}
\subcaption{$\delta = 0.01, Q=5 \times 10^{-6}$}
\end{minipage} \\
\begin{minipage}[b]{0.95\hsize}
\centering
\includegraphics[width=8cm]{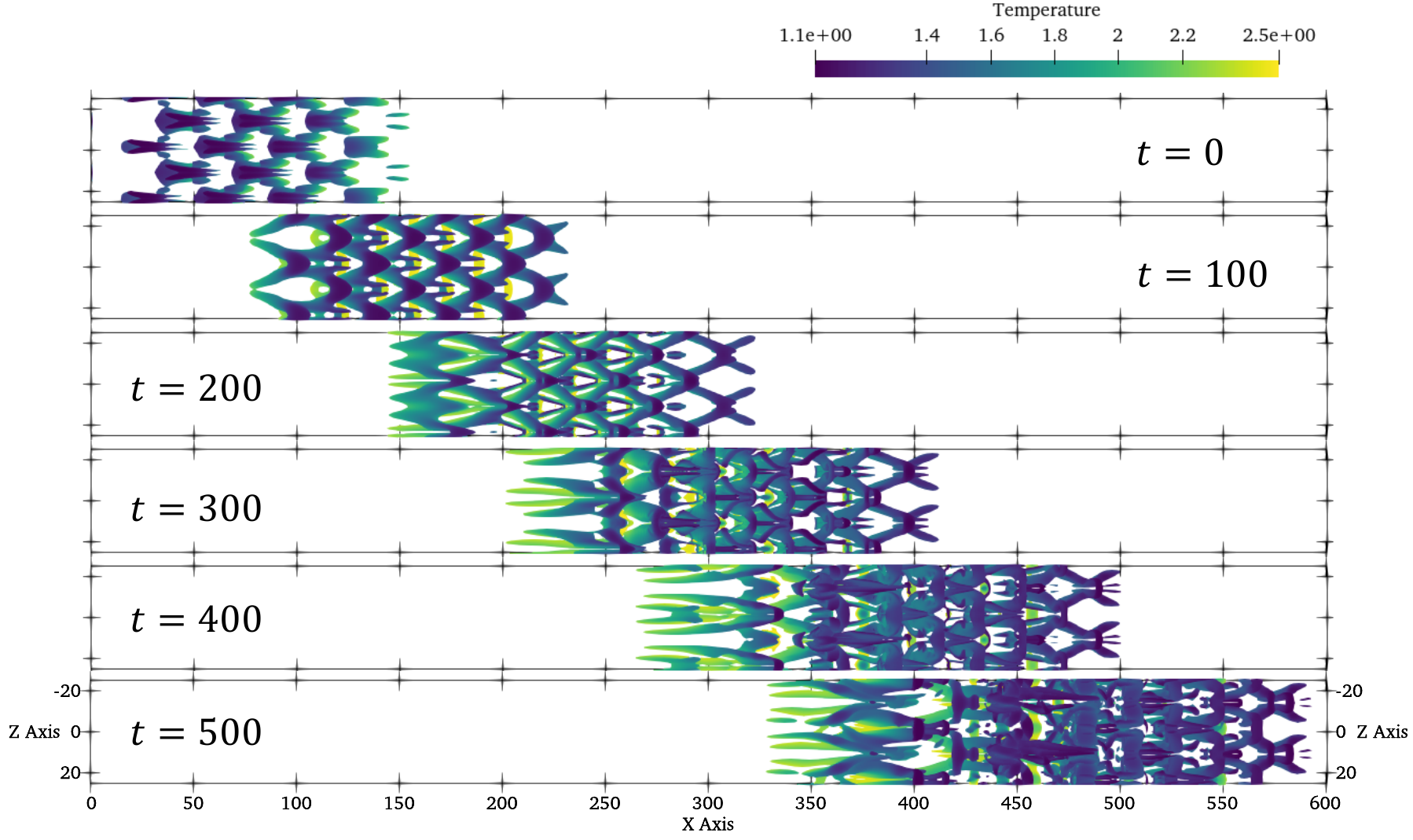}
\subcaption{$\delta = 0.05, Q = 10^{-4}$}
\end{minipage} 
\caption{
Iso-surfaces of $Q$ values of the perturbed flow field under adiabatic wall conditions ($t_f = 200$) for the determination of optimal perturbation. 
}
\label{Fig:pic_skin_d150}
\end{figure}

Figure \ref{Fig:pic_skin_d150} shows the evolution of iso-surfaces of the second invariant of the velocity gradient tensor ($Q$-criterion) for the perturbed flow field at different initial disturbance amplitudes. Fig. \ref{Fig:pic_skin_d150} (a) shows the case with an initial disturbance amplitude of $\delta = 0.01$, where the growth of disturbances up to $t = 250$ is not significantly different from the linear growth, as shown in Fig. \ref{fig:pic_dis_evol}. The main structure of the disturbance field is concentrated in the upper part of the boundary layer, forming a staggered structure. This vortical structure is similar to the transition process in oblique breakdown reported in \cite{mayer2011direct, lim2021cost}. The self-interaction mechanism of disturbance vortices results in an elongated and locally parallel vortex structure. 

\begin{figure}[ht]
\centering
\begin{minipage}[b]{0.95\hsize}
\centering
\includegraphics[width=7cm]{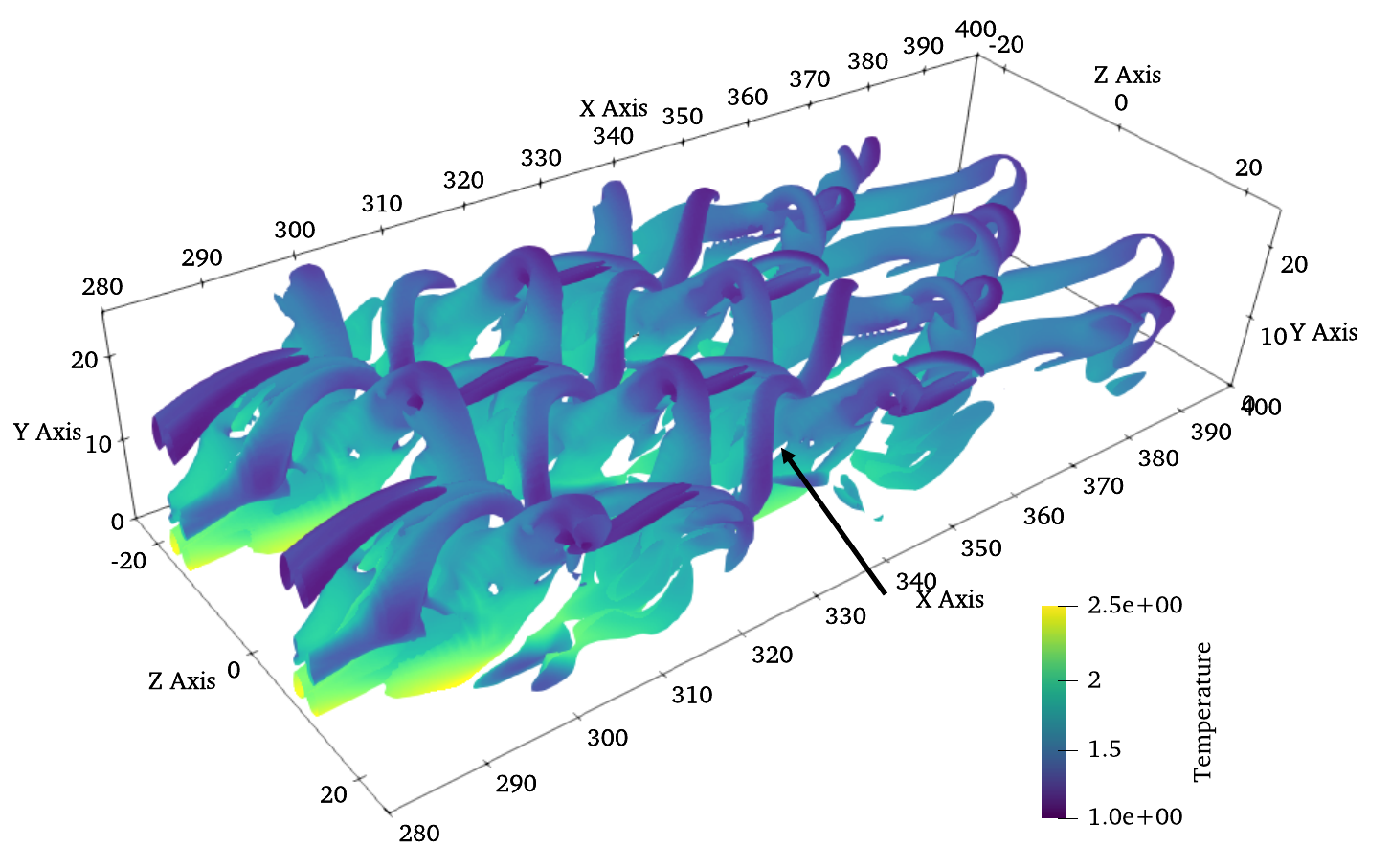}
\subcaption{$T=T_{\mathrm{ad.}}$}
\end{minipage} \\
\begin{minipage}[b]{0.95\hsize}
\centering
\includegraphics[width=7cm]{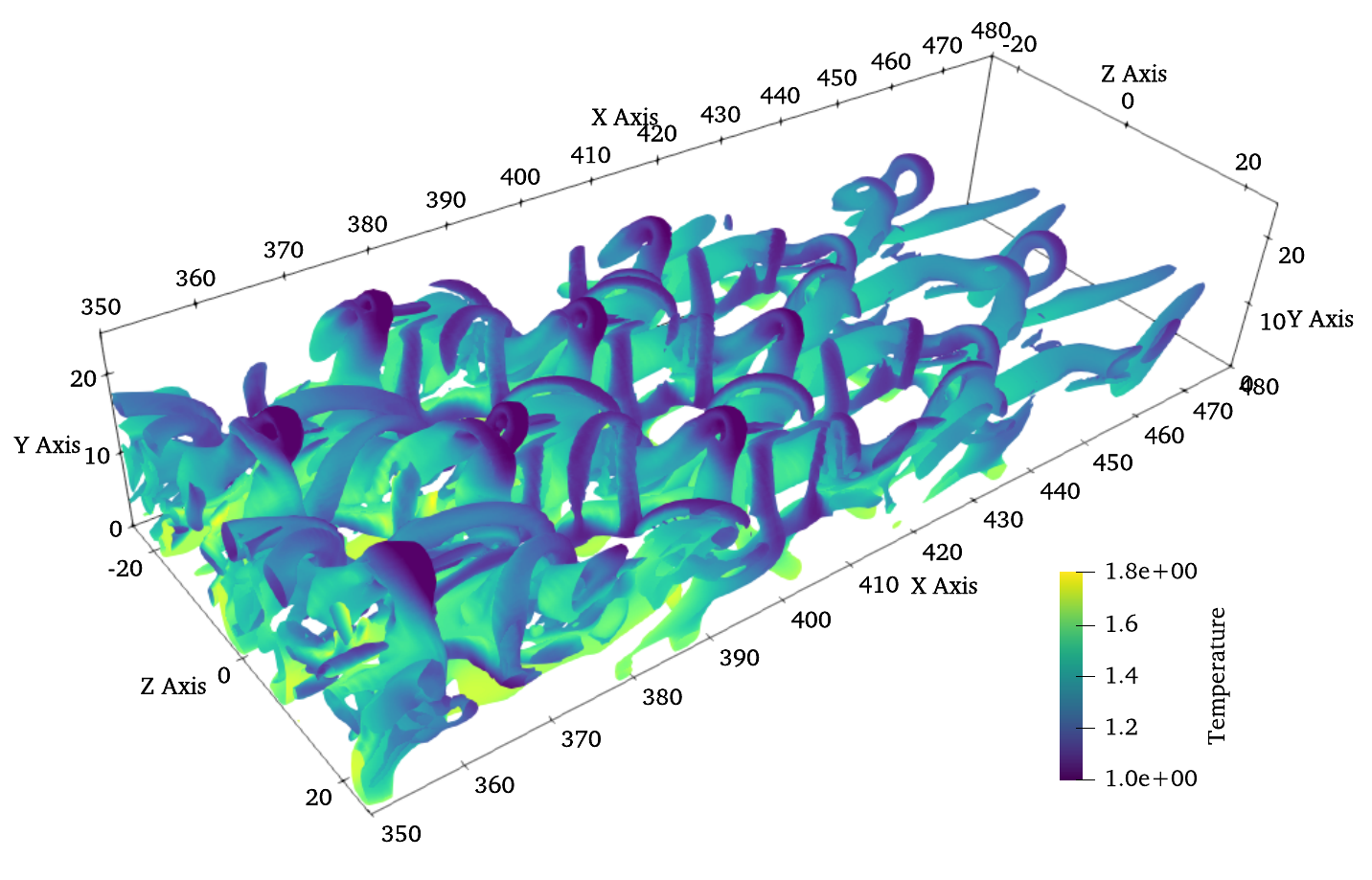}
\subcaption{$T=0.6 T_{\mathrm{ad.}}$}
\end{minipage} 
\caption{
Close-up view of $Q$-criterion iso-surfaces ($Q = 0.001$) at $t=300$, colored by temperature. 
}
\label{Fig:pic_close_vortex}
\end{figure}

For a sufficiently large initial disturbance amplitude, a three-dimensional (3D) vortex structure emerges, as shown in Fig. \ref{Fig:pic_skin_d150} (b). This figure illustrates a two-layer structure across the boundary layer thickness, which develops from an initial disturbance field that is widely distributed in the wall-normal direction. After the initial deformation of these disturbances, the upper layer organizes into a staggered arrangement of hairpin vortices accompanied by secondary spanwise vortices near the wall (see $t=100$ in Fig. \ref{Fig:pic_skin_d150} (b)). Subsequently, the interaction between the heads of the hairpin vortices and the near-wall secondary spanwise vortices generates spatially twisted vortices, as shown in Fig. \ref{Fig:pic_close_vortex} (indicated by a black arrow), whose distribution shows a similarity in S-shaped vortices~\cite{adams1996subharmonic}. Fig. \ref{Fig:pic_close_vortex} (b) shows the cooled-wall case, where similar vortex structures are observed except for differences in the degree of nonlinearity. Therefore, the thinning of the boundary layer under cooled-wall conditions promotes active vortex interactions in the wall-normal direction. 

\begin{figure}[ht] 
\centering
\includegraphics[width=12.0cm]{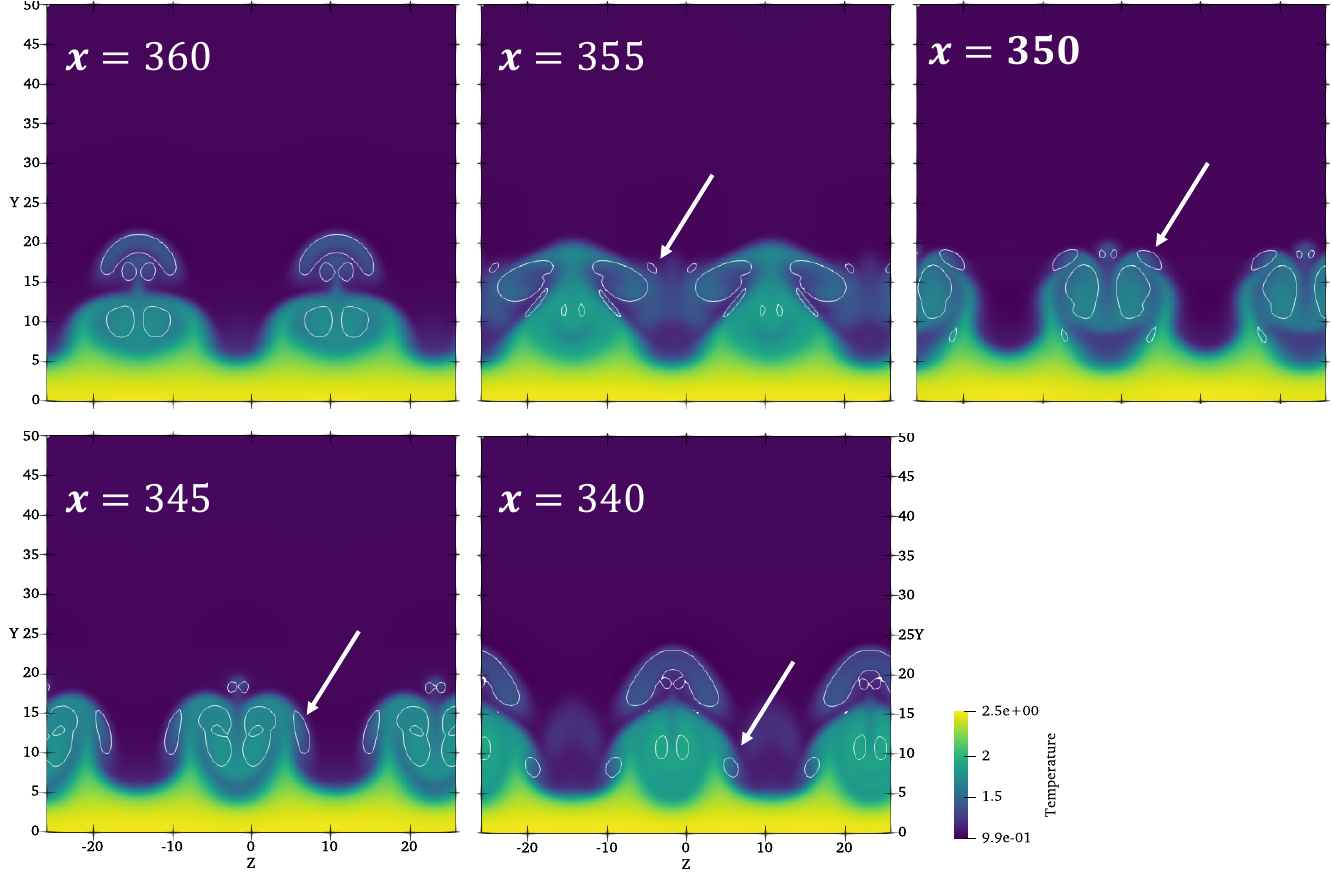} 
\caption{
Temperature distribution on cross-sectional planes at a constant streamwise locations ($x=360, 355, 350, 345, 340$) from Fig. \ref{Fig:pic_close_vortex} (a) over one period. The curves represent the $Q = 0.001$ contours and the white arrow tracks a single S-shaped vortex. 
}
\label{Fig:tem_Sline}
\end{figure}

Figure \ref{Fig:tem_Sline} shows the cross-sectional views corresponding to Fig. \ref{Fig:pic_close_vortex} (a), where the arrow tracks the structure of the S-shaped vortex. It can be seen that the streaks, which originate from the staggered vortex structure near the edge of the boundary layer, induce the transport of momentum in the wall-normal direction by the lift-up mechanism, accompanied by the temperature advection as a passive scalar. These S-shaped vortices, which are characterized by nearly wall-normal vortex filaments inclined at a steep angle relative to the streamwise direction, are distributed along the velocity shear within the streaks, as seen in Fig. \ref{Fig:pic_close_vortex} (a). This formation triggers a strong three-dimensional interaction inside the boundary layer, and the local vortex structure evolves in the wall-normal direction. During this temporal evolution, the entire structure stretches in the streamwise direction, eventually developing into a multi-scale flow structure composed of a staggered hairpin vortex structures upstream, turbulent vortices in the middle, and elongated streamwise vortices downstream. 

\section{Conclusion}

This study identifies the NLOP that triggers the transition in supersonic boundary layers at $M=3.0$. The NLOP forms staggered vortex structures near the outer edge of the compressible boundary layer, leading to the generation of S-shaped vortices and, eventually, three-dimensional structures. This time-evolution of the perturbed flow field by NLOP exhibits a distinctive evolution from that of the LOP scaled-up by a constant factor to match the same initial disturbance amplitude, which shows similar characteristics to the oblique breakdown with the secondary instability of the streaks. These differences in the transition path originates from the contraction of the streamwise wavenumber of the optimal perturbation exceeding a threshold initial disturbance amplitude between $0.01$ and $0.02$ at $t_f = 200$, where the vortex structure of LOP changed into two lines of alternating streamwise vorticity. 

Compared to the evolution of the minimal seed in the incompressible boundary layer reported by Cherubini et al. (2011)~\cite{cherubini2011minimal}, the present study shows a unique structural evolution in supersonic boundary layers. They reported that the nonlinearity drives a self-sustaining process through the interaction of streamwise vortices aligned in the spanwise direction, eventually leading to the formation of hairpin vortices and their subsequent breakdown into smaller scales. In contrast, a supersonic boundary layer is characterized by an increase in boundary layer thickness and higher relative momentum in its upper region due to the decrease in density near the wall. These characteristics highlight the vertical growth of structures during the time evolution of the minimal seed in supersonic boundary layers, which originates from the self-interaction mechanisms similar to Adams and Kleiser (1996)~\cite{adams1996subharmonic}. The characteristics of disturbance growth in the present study differ clearly from the oblique breakdown mechanism of Chang and Malik (1994)~\cite{chang1994oblique}, which is driven by a triadic interaction among a pair of oblique waves and a steady streamwise vortex. Thus, the supersonic minimal seed should be interpreted as the entire nonlinear process composed of the formation of secondary spanwise vortices and subsequent wall-normal interaction rather than a single nonlinear effect. 

Under the cooled wall condition, a comparable nonlinear self-interaction process becomes dominant at a smaller initial disturbance amplitude than under insulated conditions. Despite the suppression of linear disturbance amplification, the boundary-layer thinning induced by wall cooling promotes the generation of S-shaped vortices even at smaller initial disturbance amplitudes. These findings suggest that the vortex self-induction in supersonic boundary layers leads to local breakdown, particularly through the induction of secondary spanwise vortices near the wall. Furthermore, these results demonstrate that accounting for streamwise nonlinear interactions is essential for accurately predicting bypass transition, thereby highlighting the necessity of considering full nonlinear interactions.

\backmatter

\bmhead{Acknowledgements}

This study is supported by JST CREST JPMJCR24Q6. 

\section*{Declarations}

The authors declare no conflict of interests. The optimal perturbation datasets are available in the Zenodo repository (DOI: 10.5281/zenodo.18383903) after publication. 

\begin{appendices}

\section{Adjoint compressible N--S equation}
\label{sec:Adjoint compressible N--S equation}

The adjoint equations for compressible N--S equations are obtained in the same manner of \cite{ohmichi2021matrix} as follows: 
\begin{equation}
\partial_t \rho^* = \gamma \frac{\bar{p}}{\bar{\rho}} \bar{u}_j \partial_j p^* + \tilde{\gamma} \frac{p^*}{\bar{\rho}} \bar{u}_j \partial_j \bar{p} + \bar{u}_j \bar{u}_i \partial_j m_i^* + \frac{\bar{T}}{\bar{\rho}} S + \frac{1}{Re} \frac{\bar{u_i}}{\bar{\rho}} \partial_j \tau_{ij}^*  + \sigma(x,y) \rho^*, 
\end{equation}
\begin{equation}
\partial_t m_i^* = - \partial_i \rho^* - \bar{u}_j (\partial_j m_i^* + \partial_i m_j^*) - \gamma \frac{\bar{p}}{\bar{\rho}} \partial_i p^* - \tilde{\gamma} \frac{p^*}{\bar{\rho}} \partial_i \bar{p} - \frac{1}{\bar{\rho}} \partial_j \tau_{ij}^* + \sigma(x,y) m_i^*, 
\end{equation}
\begin{equation}
\partial_t p^* = -\bar{u}_j \partial_j p^* - \partial_i m_i^* + \tilde{\gamma} \partial_j \bar{u}_j p^* - \frac{\gamma M^2}{\bar{\rho}} S + \sigma(x,y) p^*, 
\end{equation}
where
\begin{equation}
S = \frac{1}{M^2 Re Pr} \left( - \frac{d \bar{\mu}}{d \bar{T}} \partial_i \bar{T} \partial_i p^* + \partial_i (\bar{\mu} \partial_i p^* ) \right) + \frac{1}{Re} \frac{d \bar{\mu}}{d \bar{T}} \left( - \bar{\sigma}_{ij} \partial_j m_i^* + \tilde{\gamma} \bar{\sigma}_{ij} \partial_j \bar{u}_i p^* \right) , 
\end{equation}
\begin{equation}
\tau_{ij}^* = \frac{\bar{\mu}}{Re} (\partial_j m_i^* + \partial_i m_j^* - \frac{2}{3} \delta_{ij} \partial_k m_k^*) - 2 \tilde{\gamma} p^* \bar{\tau}_{ij}. 
\end{equation}
In the above equations, $\sigma (x,y)$ are the coefficient for the external damping forces in sponge region, following \cite{bodony2006analysis}. 

The adjoint boundary condition for inflow condition $\rho'=m_i'=T'=0$ can be determined as 
\begin{equation}
m_i^* = p^* = 0
\end{equation}
and by substituting these relations to the adjoint equation and ignore the viscous term contribution, we obtain 
\begin{equation}
\nabla \rho^* + \gamma \frac{\bar{P}}{\bar{\rho}} \nabla p^*= 0. 
\end{equation}
The boundary condition for outflow condition $\nabla_n \rho' = \nabla_n m_i' = \nabla_n p'=0$, $\rho^*=m_i^*=p^*=0$ holds as an approximation for high Reynolds number. 

\section{Linear stability properties of base flow}
\label{sec:Linear stability properties of base flow}

\begin{figure}[ht]
\centering
\begin{minipage}[b]{0.45\hsize}
\centering
\includegraphics[width=5.0cm]{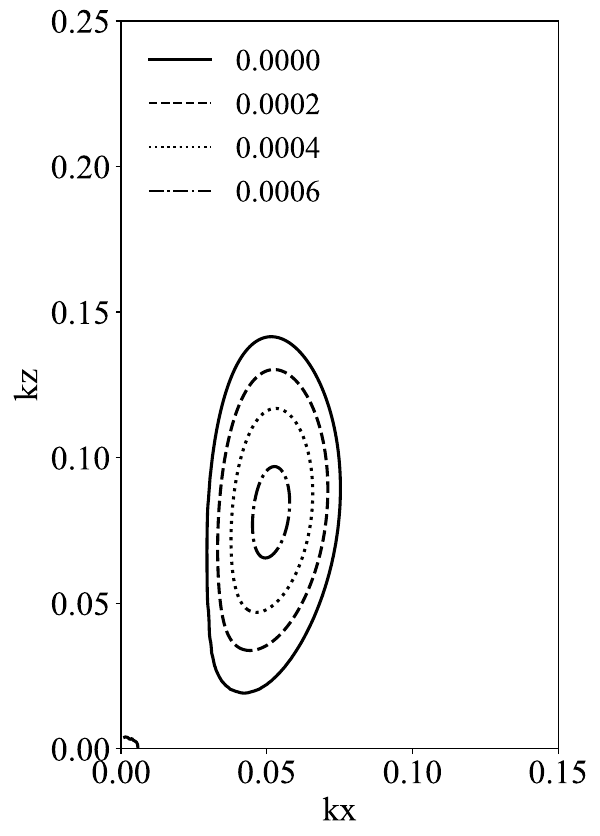}
\subcaption{$Re=300, T_w = T_{\mathrm{ad.}}$}
\end{minipage} 
\begin{minipage}[b]{0.45\hsize}
\centering
\includegraphics[width=5.0cm]{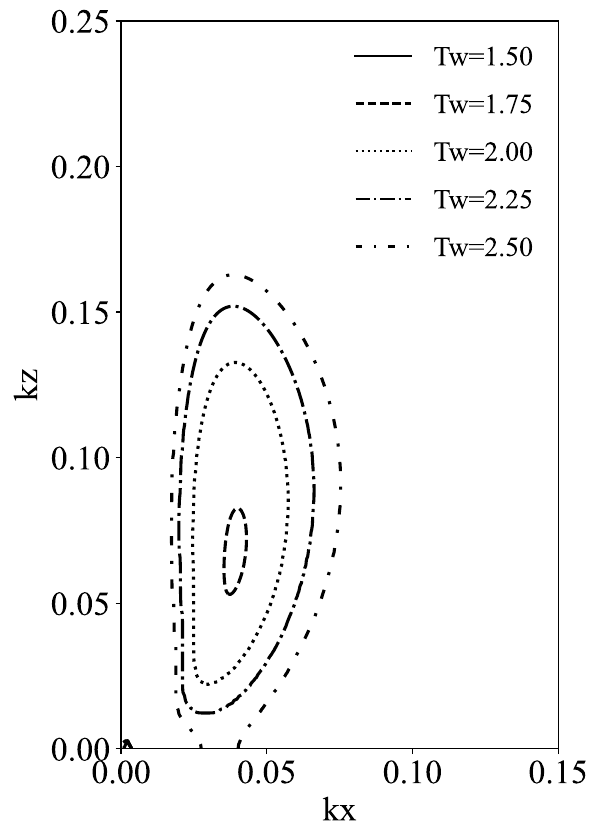}
\subcaption{$Re=500$}
\end{minipage} 
\caption{
Stability characteristics at $M=3.0$: (a) growth rate contours for the adiabatic wall condition at $Re=300$, and (b) neutral stability curves for different wall temperature conditions at $Re=500$. Note that no neutral stability curve exists for $T_w = 1.5$ in (b).  
}
\label{Fig:pic_LST_eig01}
\end{figure}

To understand the fundamental properties of the base flow field, linear stability theory (LST) is applied to both adiabatic and cooled wall temperature conditions. The implementation of LST follows the description of Malik (1990)~\cite{malik1990numerical} with excellent agreement with the reported eigenvalues therein, based on the assumption of normal modes proportional to $\exp (i (k_x x + k_z z))$, where $k_x, k_z$ are the wavenumber in streamwise and spanwise directions, respectively. The numerical discretization method is based on a sixth-order compact finite difference scheme~\cite{lele1992compact}. Based on preliminary numerical tests, the present analysis is conducted with a computational domain height of $y_{\mathrm{max}}=250$ and $256$ grid points for LST calculation. Figure \ref{Fig:pic_LST_eig01} shows a linearly unstable region for $k_z < 0.15, 0.025<k_x < 0.075$, excluding the spurious unstable region at $k_x \ll 1, k_z \ll 1$, which is significantly affected by the domain size and grid resolution. This 3D unstable region is in good agreement with the results by Hanifi et al. (1996)~\cite{hanifi1996transient} (see Fig. 7 therein). The base flow of the supersonic boundary layer is unstable to the Mack first mode, with a maximum amplification rate of $6.5 \times 10^{-4}$ at $(k_x, k_z) = (0.052, 0.08)$ and a wave angle of $57^{\circ}$. As suggested by Mack (1984)~\cite{mack1984boundary}, this linear unstable regime shrinks with decreasing surface temperature from the insulated condition, as shown in Fig. \ref{Fig:pic_LST_eig01} (b). Because the local Reynolds number at the outflow is $Re=474 (x=450)$ for non-modal analysis, the wall cooling condition of $T_w = 1.5 (=0.6 T_{\mathrm{ad.}})$ ensures subcritical conditions throughout the entire computational domain for non-modal analysis. 

\section{Grid dependency of NLOP}
\label{sec:Grid dependency of NLOP}

\begin{figure}[ht]
\centering
\begin{minipage}[b]{0.95\hsize}
\centering
\includegraphics[width=10.0cm]{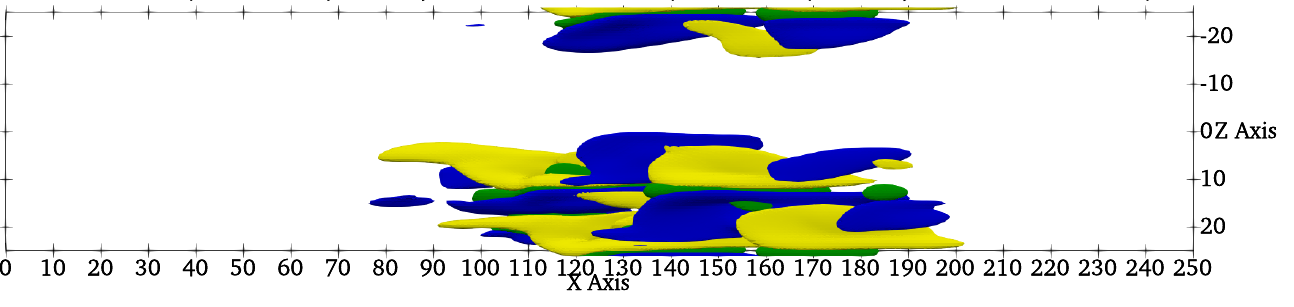}
\subcaption{Coarse Mesh}
\end{minipage} \\
\begin{minipage}[b]{0.95\hsize}
\centering
\includegraphics[width=10.0cm]{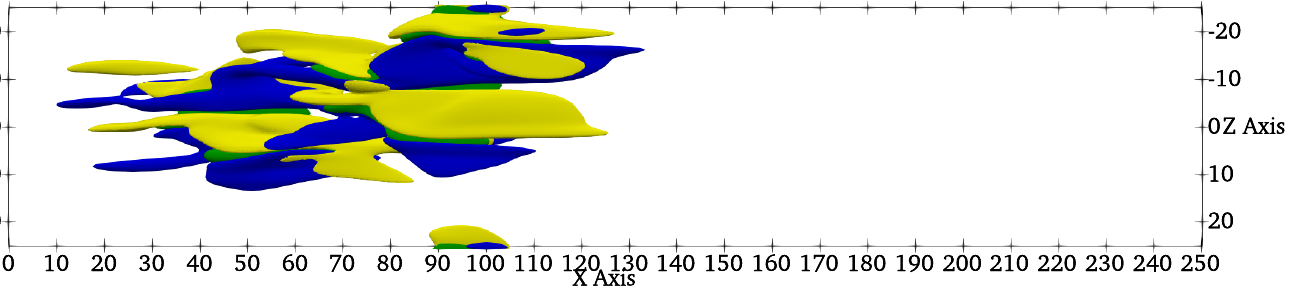}
\subcaption{Fine Mesh}
\end{minipage} 
\caption{
Spatial distribution of optimal perturbation at $t_f = 200, \delta = 0.04$. See also caption in Fig. \ref{Fig:pic_omega_pent}. 
}
\label{fig:pic_coarse_d200}
\end{figure}

This appendix illustrates the spatial grid dependence of the optimal perturbation, particularly for $\delta = 0.04$, where NLOP exhibits the strongest grid dependency on the disturbance amplification. Figure \ref{fig:pic_coarse_d200} shows the spatial distribution of the optimal perturbation computed on coarse and fine grids. Compared to the results in Fig. \ref{Fig:pic_omega_pent} using the standard grid refinement, the present case fails to capture the structural transition from a single ribbon topology (observed in $\delta = 0.02, 0.03$) to the double-ribbon topology (observed in $\delta \geq 0.04$). Instead of branching, the optimizer converges to a local maximum characterized by a single-ribbon structure, in which the streamwise distribution is adjusted to maximize the local growth rate.

\end{appendices}


\bibliography{ref_2025_boundary_M300}% common bib file

\end{document}